**Bachelor Thesis**

# A Finite Element Implementation of a Ductile Damage Model for Small Strains

*Robert Lee Gates*

robert.gates@gmail.com

Gottfried Wilhelm Leibniz University Hannover

Faculty of Civil Engineering and Geodetic Science

Institute of Mechanics and Computational Mechanics

*Univ.-Prof. Dr.-Ing. Udo Nackenhorst*
*PD Dr.-Ing. habil. Volker Berkhahn*

December 2012

**Abstract**

LEMAITRE's ductile damage model [24, 25] and a simplified variant excluding kinematic hardening were studied and implemented into computer code. For purposes of verifying the model, results from computations with the finite element method are compared to literature. It is found that the behavior expected from theory is modeled by both implementations. Quadratic levels of convergence were observed for the simplified model, while results show that convergence of the kinematic hardening implementation deteriorates with damage. It is concluded that further examination is needed to verify the correct implementation of the kinematic hardening model.

Das LEMAITRE Modell für duktilen Schaden [24, 25] und ein vereinfachender Spezialfall wurden untersucht und im Rahmen der Finite-Elemente-Methode implementiert. Um die richtige Implementierung zu gewährleisten wurden Berechnungen durchgeführt und mit Ergebnissen aus der Literatur verglichen. Hierbei zeigen beide Modelle das erwartete physikalische Verhalten. Die Konvergenz des vereinfachten Modells ist quadratisch stabil, jedoch wird deutlich, dass die Konvergenz des kinematisch härtenden Modells sich mit zunehmendem Schaden verschlechtert. Weitere Untersuchungen sind deshalb notwendig um dessen Implementierung zufriedenstellend zu verifizieren.

# Contents

# List of Figures

# List of Tables

# Notation

| | |
|---|---|
| $\mathbf{1}$ | Second order identity tensor |
| $\mathbb{1}$ | Fourth order identity tensor |
| $\boldsymbol{A}$ | Hardening thermodynamic force |
| $\boldsymbol{A} : \boldsymbol{B}$ | Double tensor contraction of two second order tensors: $\boldsymbol{A} : \boldsymbol{B} = \mathrm{tr}(\boldsymbol{A}\boldsymbol{B}^T) = \boldsymbol{A}_{ij}\boldsymbol{B}_{ij}$ |
| $\mathbb{A} : \boldsymbol{B}$ | Double tensor contraction of a fourth order tensor and a second order tensor: $(\mathbb{A} : \boldsymbol{B})_{ij} = \mathbb{A}_{ijkl}\boldsymbol{B}_{kl}$, $\quad (\boldsymbol{B} : \mathbb{A})_{kl} = \boldsymbol{B}_{ij}\mathbb{A}_{ijkl}$ |
| $\mathbb{A} : \mathbb{B}$ | Double tensor contraction of two fourth order tensors: $(\mathbb{A} : \mathbb{B})_{ijkl} = \mathbb{A}_{ijmn}\mathbb{B}_{mnkl}$ |
| $\boldsymbol{A} \otimes \boldsymbol{B}$ | Dyadic product of two second order tensors: $(\boldsymbol{A} \otimes \boldsymbol{B})_{ijkl} = \boldsymbol{A}_{ij}\boldsymbol{B}_{kl}$ |
| $\boldsymbol{A}^V$ | A tensor in Voigt notation |
| $\mathbb{C}^e$ | Isotropic elasticity tensor |
| $\boldsymbol{H}$ | Generalized hardening modulus |
| $H$ | Hardening slope |
| $\boldsymbol{N}$ | Flow vector |
| $\boldsymbol{X}$ | Kinematic hardening internal variable |
| $p$ | Hydrostatic (volumetric) stress |
| $q$ | von Mises equivalent stress |
| $\Phi$ | Yield function |
| $\Psi$ | Plastic potential function |
| $\boldsymbol{\alpha}$ | Set of internal variables associated with hardening |
| $\boldsymbol{\beta}$ | Back stress tensor and kinematic hardening thermodynamic force |
| $\boldsymbol{\varepsilon}$ | Linear strain tensor |
| $\bar{\varepsilon}^p$ | Accumulated plastic strain |
| $\boldsymbol{\eta}$ | Relative stress tensor of kinematic hardening |
| $\kappa$ | Increase in yield stress and isotropic hardening thermodynamic force |
| $\dot{\lambda}$ | Plastic multiplier or glide rate |
| $\psi$ | Helmholtz free energy potential |
| $\rho$ | Average mass density |
| $\boldsymbol{\sigma}$ | Cauchy stress tensor |
| $\sigma$ | Uniaxial stress |
| $\boldsymbol{\sigma}_d$ | Deviatoric (distortional) stress tensor |
| $\sigma_y$ | Yield stress |

# 1   Introduction

In the field of engineering, the understanding of damage mechanisms in solids is crucial to the safe operation of structures and vehicles. In order to accurately predict such mechanisms during the design stage of a product, appropriate models for the simulation of damage are required. A multitude of models currently exist in research, which can be categorized as either being based on micromechanics or on phenomenological study. A micromechanical formulation was developed by GURSON [17] and ROUSSELIER [34], accounting for ductile damage by introducing a porosity term to the yield criterion. Phenomenological damage models are based on the assumption that damage can be expressed as an internal variable, as part of the material's constitutive equations (see chapter 3). This approach was followed by LEMAITRE [24] and CHABOCHE [9], by postulating the existence of a damage dissipation potential. (cf. [7])

Unfortunately, both approaches are subject to significant limitations. As material parameters, including those used to describe damage evolution, are usually obtained from uniaxial experiments, their accuracy with respect to multi-axial states of stress is not always guaranteed. Furthermore, the damage dissipation potential, from which phenomenological models are derived, is material dependent. Modeling of different materials can therefore introduce the necessity to alter the entire model, beyond the mere identification of material parameters. In addition, most damage models are considered to be mesh-dependent (see section 4.4), while material parameters may often depend on the geometry studied. (cf. [7]) In consequence, damage mechanics is still a field of intense research.

This work aims to provide basic insight into the numerical simulation of damage within the concepts of *continuum damage mechanics*. As a model for damage in crystalline solids, LEMAITRE's phenomenological model for ductile damage [24, 25] is studied. An attempt is made to derive LEMAITRE's damage model (see chapter 4) from fundamentals on the theory of plasticity in continuum mechanics, outlined in chapter 2, followed by relevant assumptions made by the theory of continuum damage mechanics, found in chapter 3.

To accompany the rather mathematical derivation of the model, relevant physical meaning, where possible, is leant to the relations presented. In addition, underlying microstructural causes of plasticity, hardening, and damage are given in the corresponding sections. The algorithmic implementation within the finite element model is discussed in section 4.1, while a variant of LEMAITRE's damage model [11], simplified by the absence of kinematic hardening, is outlined in section 4.2. As both the kinematic hardening and simplified versions are based on the same relations outlined in chapters 2 and 3, the simplified model is not treated separately from the original. Distinctions between both models are made where appropriate.

Both the simplified and kinematic hardening versions of the model were implemented in finite element code written in MATLAB (see appendix B.1 and B.2, respectively). Difficulties encountered during the implementation are treated in section 4.3. In order to verify the correct implementation of both models, calculations with the finite element method were carried out and results were compared to reference calculations. Presentation and discussion of these results is found in section 4.4, while a short description of convergence and possible shortcomings of the kinematic hardening implementation is given in section 4.5.

# 2 Fundamentals of Plasticity

While elastic material behavior is a reversible process, plasticity describes the irreversible deformation of a solid. In metals, which are composed of irregular crystal grains, plasticity occurs by the rupture of atomic bonds and the creation, movement, and localization of dislocations. The microstructural effects leading to plasticity are briefly reviewed in chapter 3.

In the following, plasticity is treated in mathematical terms, forming the basis of material modeling in the field of continuum mechanics. The reader should be aware that most of the principal relations found in this chapter were taken from DE SOUZA NETO et al. (2008) [11], although derivation is often altered, shortened or described in further detail to aide the reader's understanding. As these can be predominantly considered to be general knowledge, explicit citations are omitted. An attempt is made to provide concise insight into the relations needed for the understanding of LEMAITRE's damage model.

As a basis for the mathematical modeling of plasticity within the regime of *infinitisemal strain theory*, the *linear strain tensor* $\boldsymbol{\varepsilon}$ can be used. Through an additive split, the elastic $\boldsymbol{\varepsilon}^e$ and plastic strain tensors $\boldsymbol{\varepsilon}^p$ are obtained:

$$\boldsymbol{\varepsilon} = \boldsymbol{\varepsilon}^e + \boldsymbol{\varepsilon}^p \, . \tag{2.1}$$

The isothermal HELMHOLTZ *free energy potential* $\psi$ can then be written as a function of the linear strain tensor, its plastic part, and a set of internal variables associated with hardening $\boldsymbol{\alpha}$, which will be discussed in section 2.3. Split into its elastic and plastic parts, the free energy potential is

$$\psi\left(\boldsymbol{\varepsilon}, \boldsymbol{\varepsilon}^p, \boldsymbol{\alpha}\right) = \psi^e\left(\boldsymbol{\varepsilon} - \boldsymbol{\varepsilon}^p\right) + \psi^p\left(\boldsymbol{\alpha}\right) \tag{2.2}$$

$$= \psi^e\left(\boldsymbol{\varepsilon}^e\right) + \psi^p\left(\boldsymbol{\alpha}\right) \, , \tag{2.3}$$

where, assuming linear and isotropic elasticity, the elastic part is given by

$$\rho \, \psi^e \left( \boldsymbol{\varepsilon}^e \right) = \frac{1}{2} \, \boldsymbol{\varepsilon}^e : \mathbb{C}^e : \boldsymbol{\varepsilon}^e \,, \tag{2.4}$$

with $\mathbb{C}^e$ being the fourth-order isotropic elasticity tensor.

The isothermal CLAUSIUS-DUHEM inequality, ensuring *thermodynamic admissibility* (dissipated energy being greater than or equal to zero), implies the state laws

$$\boldsymbol{\sigma} = \rho \frac{\partial \psi}{\partial \boldsymbol{\varepsilon}^e} = \mathbb{C}^e : \boldsymbol{\varepsilon}^e \quad \text{and} \quad \boldsymbol{\sigma} = \rho \frac{\partial \psi}{\partial \boldsymbol{\varepsilon}} = -\rho \frac{\partial \psi}{\partial \boldsymbol{\varepsilon}^p} \,, \tag{2.5}$$

while the latter shows that $-\boldsymbol{\sigma}$ is the thermodynamic force conjugate of the plastic strain (cf. [26]). With the *principle of maximum energy dissipation* it is known that of all admissible plastic states, the true plastic state maximizes the plastic *energy dissipation function*.

## 2.1 The Yield Criterion

In order to distinguish elastic from plastic deformations, a *yield criterion* is required. The associated yield function $\Phi$ can be expressed in terms of the CAUCHY stress $\boldsymbol{\sigma}$ and the *hardening thermodynamic force*

$$\boldsymbol{A} = \rho \frac{\partial \psi^p}{\partial \boldsymbol{\alpha}} \quad \text{as} \tag{2.6}$$

$$\Phi \left( \boldsymbol{\sigma}, \boldsymbol{A} \right) \leq 0 \,. \tag{2.7}$$

The boundary of the *elastic domain* is called the *yield locus*

$$\mathbb{Y} = \{ \boldsymbol{\sigma} \, | \, \Phi \left( \boldsymbol{\sigma}, \boldsymbol{A} \right) = 0 \} \tag{2.8}$$

and the restriction $\Phi = 0$ is called the yield criterion. This boundary is part of the set of *plastically admissible stresses* and represents a surface in the space of principal stresses, which is called the *yield surface*. Plastic flow only occurs on this boundary, with the yield criterion being equal to zero. When the result of the yield function is below zero, stresses are within the elastic domain

$$\mathbb{E} = \{ \boldsymbol{\sigma} \, | \, \Phi \left( \boldsymbol{\sigma}, \boldsymbol{A} \right) < 0 \} \,. \tag{2.9}$$

For a one-dimensional case of ideal plasticity, the yield function can be expressed in terms of the uniaxial stress $\sigma$ and the material's yield strength $\sigma_y$ so that

$$\Phi(\sigma) = |\sigma| - \sigma_y \leq 0\,. \tag{2.10}$$

It becomes clear that, as a result of the inequality, only stresses at or below the yield strength can occur. As a consequence, when the yield criterion is equal to zero, the material experiences plastic flow without a further increase in stress.

In general, it is assumed that plastic yielding in metals is an isochoric process and thus only dependent on *deviatoric stress* $\boldsymbol{\sigma}_d$. In contrast to *hydrostatic stress* $p$, which occurs as a result of changes in volume, deviatoric stress occurs only as a result of the distortion of a body. The CAUCHY stress tensor can be additively decomposed so that

$$\boldsymbol{\sigma} = p\mathbf{1} + \boldsymbol{\sigma}_d\,, \quad \text{where} \tag{2.11}$$

$$p = \frac{1}{3}\text{tr}(\boldsymbol{\sigma})\,. \tag{2.12}$$

Therefore, the yield criterion for metals can be expressed in terms of deviatoric stress as $\Phi(\boldsymbol{\sigma}_d, \boldsymbol{A}) = 0$.

One of the yield criteria which are appropriate for describing yielding in metals is the VON MISES ($J_2$) yield criterion. Its aim is to provide a yield criterion for multi-axial stress states using material parameters obtained from uniaxial experiments. Assuming linear elasticity, the elastic free energy can be split into hydrostatic and deviatoric components. As stated above, only the deviatoric component

$$\psi_d^e = \frac{1}{G}J_2 = -\frac{1}{2G}\text{tr}(\boldsymbol{\sigma}_d^2) = -\frac{1}{2G}\left(\boldsymbol{\sigma}_d : \boldsymbol{\sigma}_d\right) \tag{2.13}$$

has an influence on yielding. Note that, as deviators have a zero trace per definition, $J_1 = \text{tr}(\boldsymbol{\sigma}_d) = 0$, that the second invariant $J_2 = (J_1^2 - \text{tr}(\boldsymbol{\sigma}_d^2))/2$, and that $G$ is the shear modulus. When the elastic free energy of distortion reaches a critical value $\psi_d^e = \psi_{\text{crit}}$, further energy is dissipated by yielding (cf. [11]). This leads to the VON MISES yield criterion

$$\Phi(\boldsymbol{\sigma}) = q(\boldsymbol{\sigma}_d) - \sigma_y = 0\,, \tag{2.14}$$

which states that yielding occurs when a certain critical value $R(\alpha)$ of the second invariant of deviatoric stress is reached so that $J_2 = R(\alpha)$. With the VON MISES *equivalent stress*

$$q(\boldsymbol{\sigma}_d) = \sqrt{-3J_2} = \sqrt{\frac{3}{2}(\boldsymbol{\sigma}_d : \boldsymbol{\sigma}_d)} \,, \tag{2.15}$$

the uniaxial yield strength is then given by

$$\sigma_y = \sqrt{-3R(\alpha)} \,. \tag{2.16}$$

Figure 2.1: A submanifold of the VON MISES yield surface in the space of principal stresses, showing hydrostatic ($p$) and deviatoric trial stress ($\boldsymbol{\sigma}_d$).

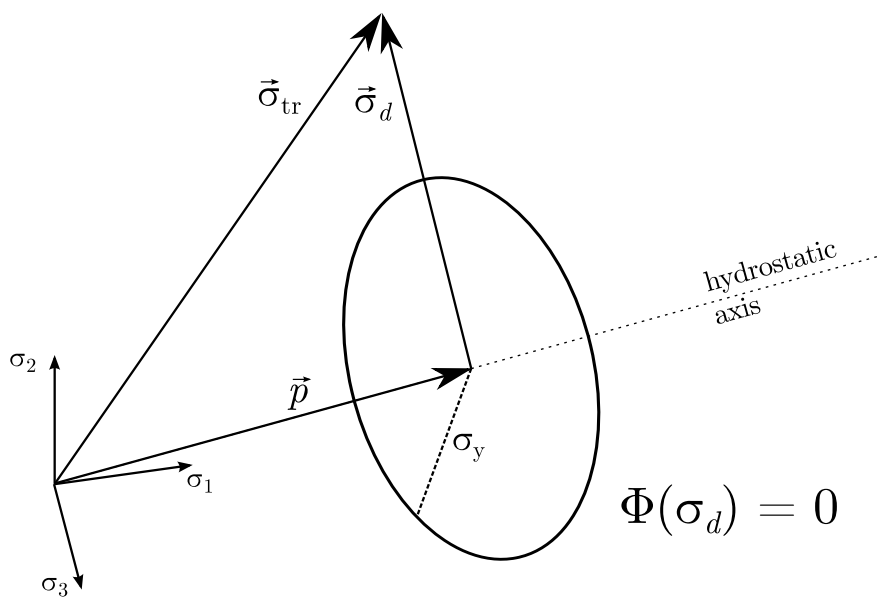

## 2.2  The Flow Rule

To describe the material's behavior within the plastic regime, the *flow rule* is introduced. It describes the evolution of the plastic strain $\boldsymbol{\varepsilon}^p$ on the boundary of the elastic domain. For a general case, the flow rule can be stated in terms of the *plastic multiplier* or *glide rate* $\dot{\lambda}$ and a generalized tensor called the *flow vector* $\boldsymbol{N}(\boldsymbol{\sigma}, \boldsymbol{A})$. The flow rule

$$\dot{\boldsymbol{\varepsilon}}^p = \dot{\lambda}\,\boldsymbol{N} \tag{2.17}$$

is restricted by the loading/unloading or KUHN-KARUSCH-TUCKER conditions:

$$\Phi \leq 0, \quad \dot{\lambda} \geq 0, \quad \Phi \dot{\lambda} = 0. \tag{2.18}$$

They result from an optimization problem of maximizing the plastic energy dissipation function for all plastically admissible states. These conditions imply that the plastic strain rate $\dot{\boldsymbol{\varepsilon}}^p$ is greater than zero when the yield criterion ($\Phi = 0$) is met. As a result, the plastic multiplier must also be greater than zero. Conversely, when the yield criterion is not met, the material is not subject to plastic straining and the plastic multiplier is zero.

The flow rule can be interpreted as a *plastic potential* function $\Psi = \Psi(\boldsymbol{\sigma}, \boldsymbol{A})$ and the flow vector (the surface normal of $\Psi$) is then defined as

$$\boldsymbol{N} = \frac{\partial \Psi}{\partial \boldsymbol{\sigma}}. \tag{2.19}$$

Generally, it should be considered that the flow potential must be a convex function of $\boldsymbol{\sigma}$ and $\boldsymbol{A}$ in the space of stresses, while being zero at the origin. This is required, as only a convex function will yield a unique state of stress for a given plastic strain rate and therefore will satisfy the CLAUSIUS-DUHEM inequality. (cf. [11])

In some models, the yield function $\Phi$ is also the flow potential $\Psi$, in that $\Phi = \Psi$. Such models are called *associative* plasticity models. In this case, the flow vector and the direction of plastic flow are normal to the yield surface. The flow vector can then be derived as follows:

$$\boldsymbol{N} = \frac{\partial \Phi}{\partial \boldsymbol{\sigma}}. \tag{2.20}$$

Considering the one-dimensional yield criterion from equation 2.10 and assuming associative plasticity, the corresponding flow rule is found according to equation 2.20 in the following way:

$$\boldsymbol{N} = \frac{\partial \Phi}{\partial \sigma} = \frac{\partial (|\sigma| - \sigma_y)}{\partial \sigma} = \frac{\partial |\sigma|}{\partial \sigma} = \frac{\sigma}{\sqrt{\sigma^2}} = \text{sign}(\sigma). \tag{2.21}$$

The PRANDTL-REUSS equations provide the associative flow vector to the isotropic VON MISES yield function from equation 2.14 as

$$\boldsymbol{N} = \frac{\partial}{\partial \boldsymbol{\sigma}} \left( \sqrt{-3J_2} \right) = \sqrt{\frac{3}{2}} \frac{\boldsymbol{\sigma}_d}{||\boldsymbol{\sigma}_d||}. \tag{2.22}$$

The tensor derivative implies the coaxiality of the flow vector and the Cauchy stress tensor (cf. [11]). This means that the principal directions of the flow vector are the same as those of the *principal stresses*.

## 2.3 Hardening Laws

In contrast to perfect plasticity, where the yield locus is constant, plastic hardening describes phenomena where the yield surface either dilates or translates under increased plastic strain. In reality, most materials exhibit both forms of hardening. Models describing such phenomena are termed as being isotropic or kinematic hardening, respectively. These concepts are discussed in further detail in sections 2.3.1 and 2.3.2.

In general, an evolution law for the hardening internal variables $\boldsymbol{\alpha}$ is required. Such an evolution law can be expressed in terms of the tensor called the generalized *hardening modulus* $\boldsymbol{H}$ and the plastic multiplier so that the *hardening law* becomes

$$\dot{\boldsymbol{\alpha}} = \dot{\lambda} \boldsymbol{H} \, . \tag{2.23}$$

As hardening only occurs in conjunction with plastic flow, the above evolution law is restricted by the optimality conditions from equation 2.18. The hardening modulus can be obtained by deriving either the plastic potential $\Psi$ or, in associative models, the yield function $\Phi$ with respect to the hardening thermodynamic force $\boldsymbol{A}$. Then the generalized hardening modulus is

$$\boldsymbol{H}(\boldsymbol{\sigma}, \boldsymbol{A}) = -\frac{\partial \Psi}{\partial \boldsymbol{A}} \left( = -\frac{\partial \Phi}{\partial \boldsymbol{A}} \right) \, . \tag{2.24}$$

### 2.3.1 Isotropic Hardening

When plastically deforming a material, dislocations increasingly nucleate. Isotropic hardening can then be assumed to be a consequence of the increased critical shear stress, as the dislocation density increases as a result of plastic flow. Although there are a multitude of theories associated with the explanation of isotropic hardening (cf. [19]), the following focuses on the theory developed by Alfred Seeger [37].

Plastic flow occurs by the movement of these dislocations in crystals and is caused by a critical shear stress, termed the Peierls and Nabarro stress. This concept was extended by Seeger [37] to account for thermal activation of screw dislocations and influence of strain rate. In an isothermal process at temperatures below the activation temperature, screw dislocations (or more generally Lomer-Cottrell junctions) are assumed to be structures of threefold symmetry and therefore to be sessile (cf. [19, 21, 37]). Moving edge dislocations are blocked by these screw dislocations, having to form sessile "jogs" for traversal of such structures. One of the components of the increase in critical shear stress is the additional energy required, as dislocations increasingly nucleate, to either transform screw dislocations into a planar and glissile form (cf. [29]) or for edge dislocations to form jogs.

Increasing dislocation density also increases the amount of pinning points for dislocations, which can contribute to an increased resistance to plastic flow. Furthermore, such pinning points are also a component required to form Frank-Read sources, which produce dislocations and lead to slip band formation (cf. [14]). For a more in-depth discussion of how this process leads to damage, refer to chapter 3.

In terms of continuum mechanics of isotropic materials, an increase in critical stress for the movement of dislocations can be described as an increase $\kappa$ in uniaxial yield stress

$$\sigma_y = \sigma_{y0} + \kappa(\overline{\varepsilon}^p) \,, \tag{2.25}$$

with $\sigma_{y0} = \text{const.}$ being the initial yield stress. This increase $\kappa(\overline{\varepsilon}^p)$ leads to a dilation of the elastic domain (see figure 2.2) and can be expressed as a function of von Mises *accumulated plastic strain*

$$\overline{\varepsilon}^p = \int\limits_0^t \sqrt{\frac{2}{3}} \, ||\dot{\varepsilon}^p|| \, \mathrm{d}t \,. \tag{2.26}$$

Therefore, the accumulated plastic strain is the hardening internal variable $\alpha = \overline{\varepsilon}^p$ for isotropic hardening and $\kappa$ is its thermodynamic force conjugate.

Due to this strain-related choice of internal variable, this kind of hardening is termed *strain hardening*, while models with work-related choices of internal variables are termed *work hardening*. In a von Mises model, both choices are equivalent (cf. [11]). Consequently, only strain hardening is further explained in the following.

Considering the rate form of the generalized yield function

$$\dot{\Phi} = \frac{\partial \Phi}{\partial \boldsymbol{\sigma}} : \dot{\boldsymbol{\sigma}} + \frac{\partial \Phi}{\partial \boldsymbol{A}} * \dot{\boldsymbol{A}} \tag{2.27}$$

and assuming $\boldsymbol{\alpha}$ to be a set of only one hardening internal variable $\overline{\varepsilon}^p$ and the only hardening thermodynamic force $\boldsymbol{A}$ to be that of isotropic hardening

$$\kappa(\overline{\varepsilon}^p) = \rho \frac{\partial \psi^p}{\partial \overline{\varepsilon}^p} \,, \tag{2.28}$$

the yield function simplifies to

$$\dot{\Phi} = \frac{\partial \Phi}{\partial \boldsymbol{\sigma}} : \dot{\boldsymbol{\sigma}} + \frac{\partial \Phi}{\partial \kappa} \, \dot{\kappa} = \frac{\partial \Phi}{\partial \boldsymbol{\sigma}} : \dot{\boldsymbol{\sigma}} + \frac{\partial \Phi}{\partial \overline{\varepsilon}^p} \, \dot{\overline{\varepsilon}}^p \,. \tag{2.29}$$

From the optimality conditions it follows that, when plastic flow occurs, $\dot{\Phi} = 0$. For a VON MISES yield function $\Phi$, with its PRANDTL-REUSS flow vector $\boldsymbol{N} = \partial \Phi / \partial \boldsymbol{\sigma}$ from equation 2.22, the yield stress from equation 2.25, and the hardening slope

$$H(\overline{\varepsilon}^p) = -\frac{\partial \Phi}{\partial \overline{\varepsilon}^p} = \frac{\partial \kappa}{\partial \overline{\varepsilon}^p} \,, \tag{2.30}$$

the following relation is obtained:

$$\dot{\Phi} = \boldsymbol{N} : \dot{\boldsymbol{\sigma}} - H \, \dot{\overline{\varepsilon}}^p \,. \tag{2.31}$$

As this yield function, by associativity, can be assumed to be a flow potential, the generalized hardening modulus is found according to equation 2.24 as follows:

$$\boldsymbol{H} = -\frac{\partial \Phi}{\partial \kappa} = -\frac{\partial(-\kappa)}{\partial \kappa} = 1 \,. \tag{2.32}$$

Therefore, the evolution law for the accumulated plastic strain in associative models is simply

$$\dot{\overline{\varepsilon}}^p = \dot{\lambda} \,. \tag{2.33}$$

## 2.3.2   Kinematic Hardening

When materials, especially polycrystalline metals, are loaded and hardened in one direction, the stress resistance in the opposite direction is decreased, leading to a translation of the yield surface. This phenomenon is called the BAUSCHINGER effect and is assumed by OROWAN [31] to be caused by the localization of mobile dislocations at precipitates and grain boundaries after initial hardening. These "obstacles" keep the dislocations from moving further under continued tension, causing a local self-equilibrated *back-stress*. Under reversed loading, this back-stress contributes to dislocation motion in the reverse direction and the yield stress is reduced. (cf. [1, 31])

The translation of the yield surface is modeled by *kinematic hardening*, where the deviatoric stress tensor $\boldsymbol{\sigma}_d$ is reduced by the back-stress tensor $\boldsymbol{\beta}$. This difference is expressed as the *relative stress* tensor

$$\boldsymbol{\eta}(\boldsymbol{\sigma}, \boldsymbol{\beta}) \equiv \boldsymbol{\sigma}_d - \boldsymbol{\beta}\,. \tag{2.34}$$

Kinematic hardening can be introduced to a plasticity model by simply replacing the CAUCHY stress with the relative stress. Thus, the translation of the yield surface is merely the back-stress tensor, which is also the thermodynamic hardening force $\boldsymbol{A}$ for kinematic hardening. For a VON MISES yield surface the yield function is then expressed as

$$\Phi(\boldsymbol{\sigma}, \boldsymbol{\beta}) = \sqrt{-3J_2(\boldsymbol{\eta})} - \sigma_y\,. \tag{2.35}$$

Mind that the yield surface only undergoes translation on the deviatoric plane. As a consequence, the back stress and the relative stress are deviatoric. The flow vector, by associativity, is then simply the PRANDTL-REUSS flow vector of the following form:

$$\boldsymbol{N} = \sqrt{\frac{3}{2}}\,\frac{\boldsymbol{\eta}}{||\boldsymbol{\eta}||}\,. \tag{2.36}$$

The thermodynamic conjugate to the back-stress tensor is the second-order tensor hardening internal variable $\boldsymbol{X}$ (or back-strain tensor, cf. [26]). It is found analogous to equation 2.6 so that

$$\boldsymbol{X} = \rho\frac{\partial\psi^p}{\partial\boldsymbol{\beta}}\,. \tag{2.37}$$

The derivation of the generalized hardening modulus $\boldsymbol{H}$ and therefore the derivation of the evolution equation for $\boldsymbol{X}$ is accomplished by first assuming the plastic free energy to be that of the SMALLCAPS{Armstrong-Frederick} kinematic hardening law:

$$\rho\psi^p = \frac{1}{3}H\left(\boldsymbol{X} : \boldsymbol{X}\right). \tag{2.38}$$

The hardening internal variable then is

$$\boldsymbol{X} = \frac{3}{2H}\,\boldsymbol{\beta}\,. \tag{2.39}$$

Substituting equation 2.39 into equation 2.38 yields:

$$\rho\psi^p = \frac{3}{4H}\left(\boldsymbol{\beta} : \boldsymbol{\beta}\right). \tag{2.40}$$

A flow potential of the form $\Psi = \Phi + \psi^p$ is assumed. Let $a = (2/3)H$ and $b = 1/\rho$, then the flow potential is

$$\Psi = \Phi + \frac{b}{2a}(\boldsymbol{\beta} : \boldsymbol{\beta})\,. \tag{2.41}$$

In accordance with equation 2.24, the generalized hardening modulus is found to be

$$\boldsymbol{H} = -\frac{\partial\Psi}{\partial\boldsymbol{\beta}} = -\frac{\partial\Phi}{\partial\boldsymbol{\beta}} - \frac{b}{2a}\frac{\partial(\boldsymbol{\beta} : \boldsymbol{\beta})}{\partial\boldsymbol{\beta}} \tag{2.42}$$

$$= \sqrt{\frac{3}{2}}\,\frac{\boldsymbol{\eta}}{||\boldsymbol{\eta}||} - \frac{b}{a}\,\boldsymbol{\beta}\,. \tag{2.43}$$

Consequently, the evolution law for the hardening internal variable is

$$\dot{\boldsymbol{X}} = \dot{\lambda}(\boldsymbol{N} - \frac{b}{a}\,\boldsymbol{\beta}) = \dot{\boldsymbol{\varepsilon}}^p - \dot{\lambda}\,\frac{b}{a}\,\boldsymbol{\beta}\,. \tag{2.44}$$

The above equation is an extension of the SMALLCAPS{Prager} kinematic hardening law called the SMALLCAPS{Armstrong-Frederick} kinematic hardening law. The plastic free energy potential from equation 2.38 introduces the effect of back-stress saturation to the SMALLCAPS{Prager} kinematic hardening law and is represented by the above equation's second term. The evolution of back-stress above a maximum limit value of $||\boldsymbol{\beta}||$ is zero, the material then behaves as perfectly plastic. (cf. [11])

Figure 2.2: Dilation of the yield surface as a consequence of isotropic hardening, showing the hardening thermodynamic force ($\kappa$).

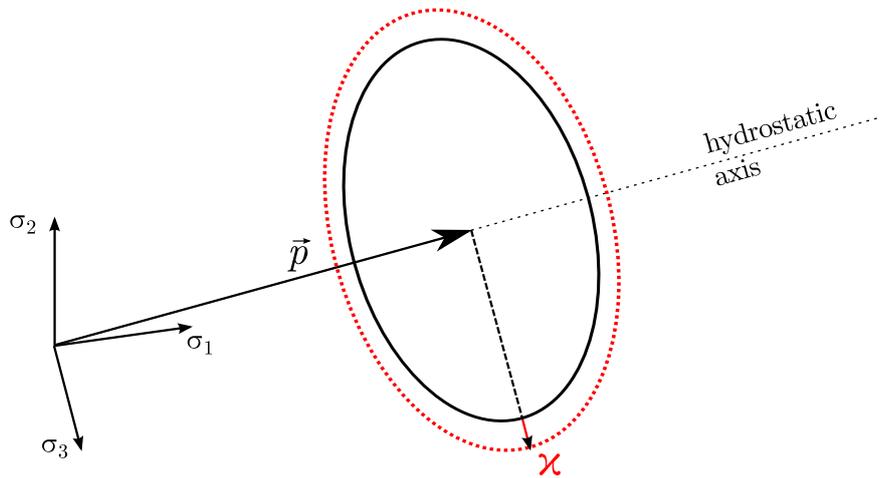

Figure 2.3: Deviatoric translation of the yield surface in kinematic hardening, showing back stress ($\boldsymbol{\beta}$) and relative trial stress ($\boldsymbol{\eta}$).

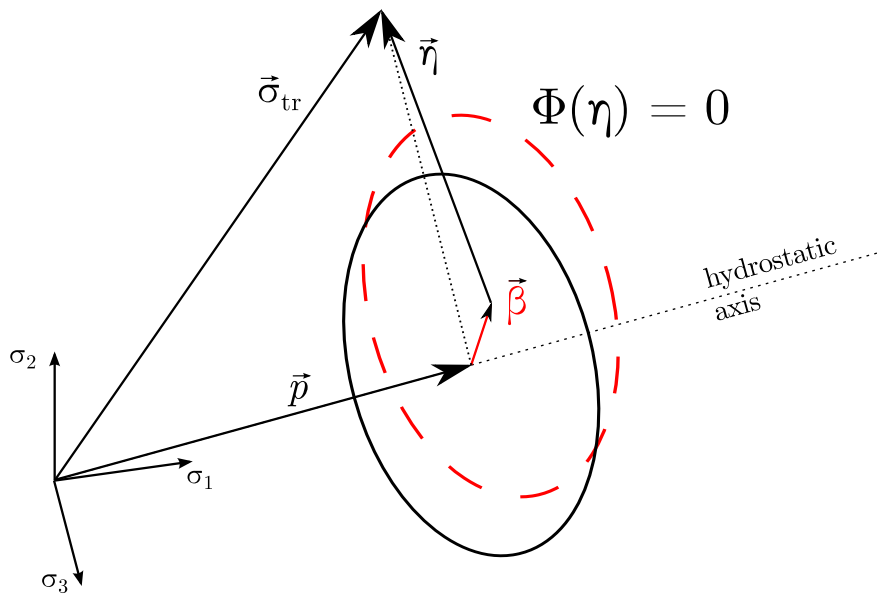

# 3 Fundamentals of Continuum Damage Mechanics

To fully describe a material's behavior after yielding, the material's inner deterioration and the process leading to its ultimate failure must be taken into account. The study of *continuum damage mechanics* attempts to describe damage in terms of a continuous field. This effectively results in interpreting the average size and density of cracks within a given infinitesimal volume. In contrast to this continuous formulation, *fracture mechanics* attempts to study the effect of the discontinuities represented by cracks on the surrounding continuum.

Within the nonlinear finite element method, both fields of study can be represented. The continuous approach can be implemented without further alterations to the method, while discontinuous methods require either the remeshing of the body or the application of techniques such as the XFEM [28]. This chapter aims to provide the reader some brief insight into the mechanisms of damage within the realm of materials science, as well as to define appropriate continuous measures of damage.

## 3.1 Mechanisms of Damage in Crystalline Solids

At the microscale level, damage is a result of dislocation dynamics. Acknowledge that crystalline solids exhibit microscopic crystallographic-texture-based anisotropy (cf. [19]). Thus, the critical shear stress, which can also be interpreted as a resistance force to dislocation motion, and the microstress are not evenly distributed throughout the material (cf. [36]). Microstress usually localizes at crystal defects, grain boundaries, and material interfaces (cf. [26]), while minimum critical shear stress is present at lattice planes characterized by the shortest Burgers vector and therefore possessing the highest atomic density (cf. [19]).

When microstress reaches the *critical shear stress* or PEIERLS-NABARRO stress locally, dislocation motion initiates. This motion is also termed *dislocation glide* or *slip* in materials science, leading to plastic deformation of the crystal lattice and ultimately of the material as a whole. As a consequence of the reduced critical shear stress on closely-packed lattice planes, such planes often exhibit a behavior called *planar slip*. Planar slip is characterized by the movement of whole lattice planes in the direction of highest atomic density as a result of dislocation glide. (cf. [19])

Frequently, a moving dislocation encounters irregularities in the crystal lattice, which can serve as pinning points for these dislocations. Such pinning points are the basis for the production of new dislocations via FRANK-READ sources (cf. [14]). These sources of *dislocation multiplication* lead to the localization of dislocations and ultimately to the formation of well-spaced stepped structures of regular length termed *slip bands*, which were first discovered in 1903 by EWING and HUMFREY [13]. The forming of such structures presents the primary mechanism for fatigue crack initiation and growth, which will be discussed briefly later. It should be noted that their spacing is given by a regular multiple of the BURGERS vector (cf. [19]) and their length is determined by the dissipation equilibrium of the FRANK-READ source (cf. [14]).

While the nucleation of dislocations can already be seen as damage, on the mesoscale, damage is represented by *microvoid* nucleation and *coalescence* (cf. [26]). *Microvoid* nucleation is believed to occur as a result of interface decohesion, e.g. at inclusions or precipitates, slip band intersection, and particle cracking (cf. [15, 19]). Consider an arbitrary *representative volume element* (RVE, see figure 3.2) embedded in the loaded material. Within this RVE, microvoids form by the aforementioned mechanisms and, as damage progresses, these microvoids grow, eventually coalescing to form large cracks (cf. [19, 26]). Microvoid growth is believed to be a consequence of planar slip, as experimental observation shows that microvoid walls have been found to exhibit wavy markings, a result of wavy or serpentine glide of slip planes (cf. [15]). A possible slip model for this process proposed by NAGPAL et al. (1973) [30] is shown in figure 3.1. At the macroscale level, the failure of an RVE can be regarded as the initiation of a *technical crack*. Further crack growth is studied by the field of fracture mechanics.

In general, materials can be classified as being either *brittle* or *ductile*. From a standpoint of materials science, the measure of ductility is given by dislocation mobility (cf. [35]). One factor influencing dislocation mobility is the so-called *dislocation width*, being a measure of lattice distortion surrounding a dislocation. For a further explanation of dislocation width, the reader is referred to HERTZBERG (1996) [19]. From the definition of PEIERLS-NABARRO stress, it is found that critical shear stress and dislocation width are inversely related. It can be assumed that, among other factors, a ductile material must exhibit a relatively low critical shear stress, allowing ductile deformation as a result of plastic flow at relatively low stresses. In consequence, ductile materials must have a relatively large dislocation width.

From these explanations, it can be inferred that the size of the *plastic zone* surrounding a given crack tip is directly related to dislocation mobility, critical shear stress, and therefore ductility. Thus, brittle cracking exhibits a relatively small extent of the plastic zone and, consequently, negligible plastic deformation as a prelude to fracture. In turn, crack propagation within a ductile material is characterized by an enlarged plastic zone as well as obviously noticeable plastic deformation as a result of relatively low critical shear stresses and high dislocation mobility.

In addition to brittle and ductile damage, another notable mechanism is *creep* damage. As briefly mentioned in section 2.3.1, dislocation mobility is affected by thermal activation. At high temperatures above the activation temperature, screw dislocations are transformed to their glissile form (cf. [29]). This can lead to a predominantly glissile structure without hardening effects and therefore to increasing plastic deformation and microvoid nucleation at constant levels of stress.

Finally, a process by which failure occurs far below the yield stress of the material under alternating loads is called *fatigue*. The latter is characterized by slip band formation at multiple sites, usually at the surface or at inclusions, and subsequent microcrack initiation. Under cyclic loading, such microcracks combine to form small radial cracks, eventually coalescing to form the fatigue macrocrack, which then cyclically propagates throughout the material until conventional fracture occurs. (cf. [33])

Crack propagation in fatigue is characterized by visible striations found on the crack surfaces (cf. [27]) as well as the area normal of the crack surface being parallel to the direction of maximum tensile stress. The latter is a consequence of the slanted orientation of slip bands at both sides of the crack tip (cf. [33]). This is an important distinction from ductile fracture in plane stress states, where the fracture surface twists into a plane of maximum shear. In general, planes of maximum shear are oriented at a 45° angle between maximum and minimum tensile stresses. Near the surface of a specimen, the material is in a state of plane stress, the minimum tensile stress thus being parallel to the surface's area normal. Consequently, the crack twists into a plane at an orientation of 45° to the surface. Such areas of slanted crack surfaces are termed *shear lips* and are a clear indicator of ductile fracture.

## 3.2   Scalar Damage Variables

As the explanations of the previous section suggest, in general, damage is considered to be anisotropic, depending on the orientation of dislocation slip within the lattice as well as the spatial orientation of microvoids. Furthermore, especially under cyclic loading in fatigue problems, crack closure effects occur either as a result of the *nominal stress* being compressive or, in tension, as a result of environment-induced crack tip oxidation reactions. Crack closure can be modeled by considering a tensile-compressive split of the Cauchy stress tensor, while anisotropy requires the definition of higher order damage variables (cf. [11,22]).

For simplicity, isotropy is assumed in the following. This can be interpreted in terms of materials science as the microvoid cross-section being identical regardless of orientation, effectively reducing an ellipsoidal void to a spherical one. As a consequence of this assumption, the damage variable can be reduced to a scalar. To describe creep failure, such a scalar damage variable was first introduced by Kachanov [20] and later given physical meaning by Rabotnov [32]. It was assumed that damage can be expressed by a reduction in cross-sectional area. (cf. [11]) With the load-bearing area of the undamaged state $A_0$ and of the damaged state $A$, the scalar damage variable $D$ is then defined as

$$D = \frac{A_0 - A}{A_0} \in [0, 1], \quad A_0 > A.$$ (3.1)

A ruptured material corresponds to a cross-sectional area $A = 0$, and therefore the damage variable takes on a value of $D = 1$. In order to derive damage-induced strain-rate variations, a so-called *effective stress* $\sigma_{\text{eff}}$ was defined as a function of *true stress* $\sigma$ and the damage variable $D$ (cf. [11]):

$$\sigma_{\text{eff}} = \frac{\sigma}{1 - D}.$$ (3.2)

It should be noted that true stress is the force per cross-sectional area in the undamaged state, while effective stress is the force per cross-sectional area of the damaged state. Thus, in mathematical terms, as damage progresses, true stress remains constant under an invariable load, while effective stress approaches infinity.

As cross-sectional area is only weakly defined and proves difficult to measure, a dam-

Figure 3.1: Coalescence of microvoids via slip systems from HANCOCK and MACKENZIE (1976) [18], as proposed by NAGPAL et al. (1973) [30].

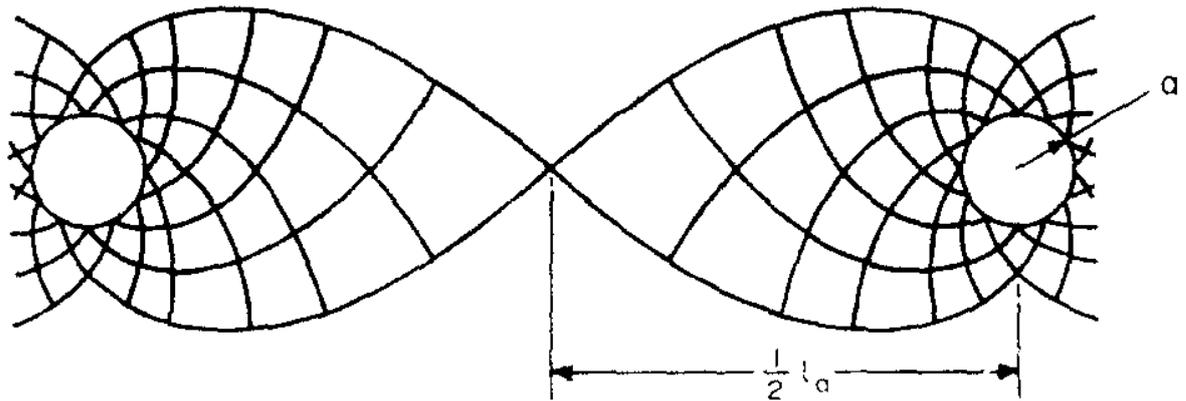

Figure 3.2: Ductile damage within a representative volume element from DE SOUZA NETO et al. (2008) [11].

virgin material          nucleation of microscopic          growth, coalescence
                            cracks and voids                and macroscopic fracturing

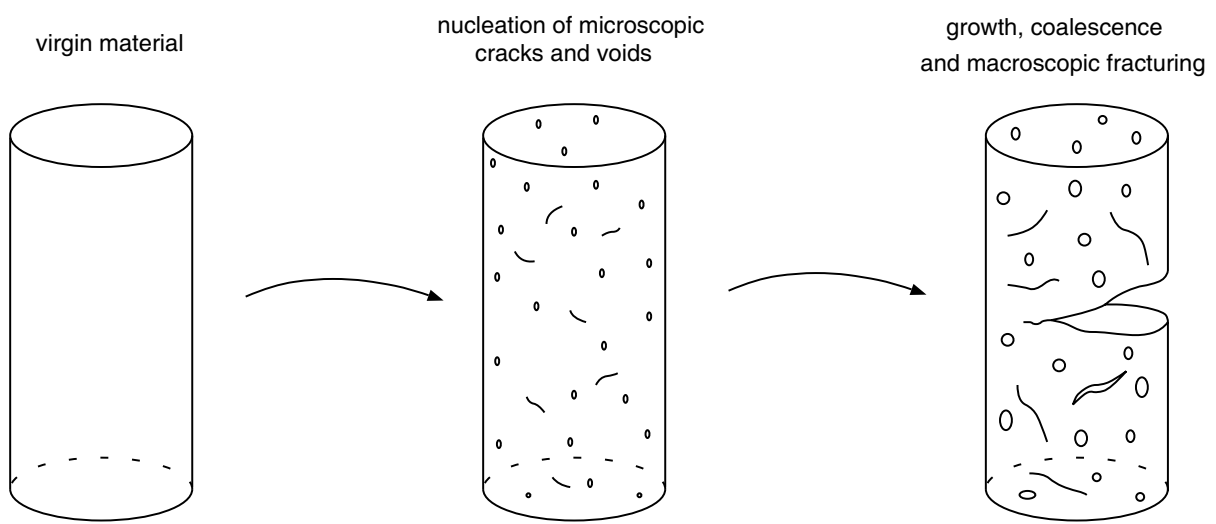

age variable based on Young's modulus was proposed by LEMAITRE [23]. This author's *hypothesis of strain equivalence* states that the constitutive laws describing the deformation behavior of a damaged material are the same as those of a virgin material with the true stress replaced by the effective stress. (cf. [11]) Taking this into account, a damaged material's stress-strain law is HOOKE's law for the undamaged material ($\sigma = E_0 \, \varepsilon^e$), expressed in terms of its YOUNG's modulus $E_0$, with the true stress replaced by the effective stress. Hence, for a one-dimensional case

$$\sigma_{\mathrm{eff}} = E_0 \, \varepsilon^e \, , \tag{3.3}$$

where $E_0$ is a material constant. For a given external load and cross-sectional area, true stress is constant by definition. In order to achieve strain equivalence, a YOUNG's modulus $E$ describing the damaged configuration is introduced by LEMAITRE so that the true stress becomes

$$\sigma = E \, \varepsilon^e \, . \tag{3.4}$$

Substituting this strain-equivalent true stress into equation 3.2 and the result thereof into equation 3.3 finally yields the relation between the Young's moduli for damaged ($E$) and virgin ($E_0$) materials:

$$E = (1 - D)E_0 \, . \tag{3.5}$$

Therefore, KACHANOV's damage variable $D$ from equation 3.1 is redefined by LEMAITRE as follows:

$$D = \frac{E_0 - E}{E_0} \in [0, 1] \, , \quad E_0 > E \, . \tag{3.6}$$

As damage progresses, the original elastic modulus is reduced, leading to a decrease in stiffness and, finally, loss of load-bearing capacity.

# 4 Lemaitre's Damage Model

In order to describe isotropic ductile plastic damage within the theory of continuous macroscopic elastoplasticity, a set of constitutive equations was proposed by JEAN LEMAITRE [24, 25]. These constitutive equations are found by consistently applying the hypothesis of strain equivalence to the laws given in chapter 2. Again, it should be noted at this point that some of the facts and many principal equations in this chapter were found in the work of DE SOUZA NETO and coworkers (2008) [11]. The specific algorithm followed was also obtained from the latter and original literature. In the following, only the original authors are cited.

For the elastic contribution to the free energy given by equation 2.4, the application of the hypothesis of strain equivalence yields

$$\rho \, \psi^{ed} \left( \boldsymbol{\varepsilon}^e, D \right) = \frac{1}{2} \, \boldsymbol{\varepsilon}^e : (1 - D) \mathbb{C}^e : \boldsymbol{\varepsilon}^e \,, \tag{4.1}$$

which is termed the *elastic-damage potential*. Consequently, the elastic law is given by

$$\boldsymbol{\sigma}_{\text{eff}} = \mathbb{C}^e : \boldsymbol{\varepsilon}^e \,, \tag{4.2}$$

or, in terms of true stress, by

$$\boldsymbol{\sigma} = (1 - D) \mathbb{C}^e : \boldsymbol{\varepsilon}^e \,. \tag{4.3}$$

The thermodynamic force conjugate to the damage internal variable is then found by differentiation to be

$$Y = \rho \, \frac{\partial \psi^{ed}}{\partial D} = -\frac{1}{2} \, \boldsymbol{\varepsilon}^e : \mathbb{C}^e : \boldsymbol{\varepsilon}^e \,, \tag{4.4}$$

and is termed the *damage strain energy release rate*. Note that $-Y$ is the continuum mechanics equivalent of the damage strain energy release rate of fracture mechanics $G$ (cf [8, 24]).

The plastic contribution to the free energy is the sum of isotropic and kinematic hardening contributions. For ARMSTRONG-FREDERICK kinematic hardening, this is given by equation 2.38. The isotropic hardening contribution follows from equation 2.28 and is simply the integral of the isotropic hardening force $\kappa(R)$, given by an arbitrary, experimentally determined function, with respect to the associated internal variable $R$. Therefore, the plastic contribution to the free energy can be written as:

$$\rho \, \psi^p \, (R, \boldsymbol{X}) = \int \kappa(R) \, \mathrm{d}R + \frac{a}{2}(\boldsymbol{X} : \boldsymbol{X}) \, . \tag{4.5}$$

From equation 2.37 it is known that

$$\boldsymbol{\beta} = \rho \frac{\partial \psi^p}{\partial \boldsymbol{X}} = a \boldsymbol{X} \, , \tag{4.6}$$

and thus, in terms of $\boldsymbol{\beta}$,

$$\rho \, \psi^p \, (R, \boldsymbol{\beta}) = \int \kappa(R) \, \mathrm{d}R + \frac{1}{2a}(\boldsymbol{\beta} : \boldsymbol{\beta}) \, . \tag{4.7}$$

By applying the hypothesis of strain equivalence to the VON MISES yield function given by equation 2.35 and by including isotropic hardening, the following yield function for the LEMAITRE model is obtained (cf. [26]):

$$\Phi(\boldsymbol{\sigma}, \boldsymbol{\beta}, \kappa, D) = \sqrt{-3J_2 \left( \frac{\boldsymbol{\sigma}}{1-D} - \boldsymbol{\beta} \right)} - \sigma_y - \kappa(R) \, . \tag{4.8}$$

Although the above is the formal definition of the yield function as outlined by LEMAITRE (1996) [26], DE SOUZA NETO et al. [10, 11] define the yield function as

$$\Phi(\boldsymbol{\sigma}, \boldsymbol{\beta}, \kappa, D) = \frac{\sqrt{-3J_2(\boldsymbol{\eta})}}{1-D} - \sigma_y - \kappa(R) \, . \tag{4.9}$$

It appears that, in the latter definition, $\boldsymbol{\beta}$ is considered to be some sort of effective stress. If this assumption is correct, it would violate the experimental consequence that damage equally reduces yield stress, the isotropic hardening thermodynamic force, as well as the back-stress (cf. [26]). This is graphically represented by a stress-strain curve in figure 4.1. As the algorithmic implementation outlined in the following is largely based on the works of DE SOUZA NETO et al., the yield function from equation 4.9 is adopted.

Extending equation 2.41 by a term related to damage as a power function of $-Y$

(cf. [11, 25]) yields the convex (for $a, b, r, s \in \mathbb{R}^+$; cf. [10]) flow potential

$$\Psi = \Phi + \frac{b}{2a}(\boldsymbol{\beta} : \boldsymbol{\beta}) + \frac{r}{(1-D)(s+1)} \left(\frac{-Y}{r}\right)^{s+1} . \tag{4.10}$$

For phenomena independent of time, the evolution law for the damage internal variable (cf. [25]) is then given by

$$\dot{D} = -\dot{\lambda}\frac{\partial \Psi}{\partial Y} = \frac{\dot{\lambda}}{1-D} \left(\frac{-Y}{r}\right)^s . \tag{4.11}$$

Recall equation 2.39 and consequently that the rate of the internal variable related to kinematic hardening $\dot{\boldsymbol{X}} = \frac{1}{a}\dot{\boldsymbol{\beta}}$. The evolution law for the back-stress is then found by substituting the latter relation into equation 2.44 to be

$$\dot{\boldsymbol{\beta}} = \dot{\lambda}(a\boldsymbol{N} - b\boldsymbol{\beta}) . \tag{4.12}$$

Figure 4.1: Stress-strain curve after yielding from LEMAITRE (1996) [26]. Note that damage occurs once the damage threshold ($\varepsilon_{pD}$) is reached, that $X = \beta$, and that $R = \kappa$.

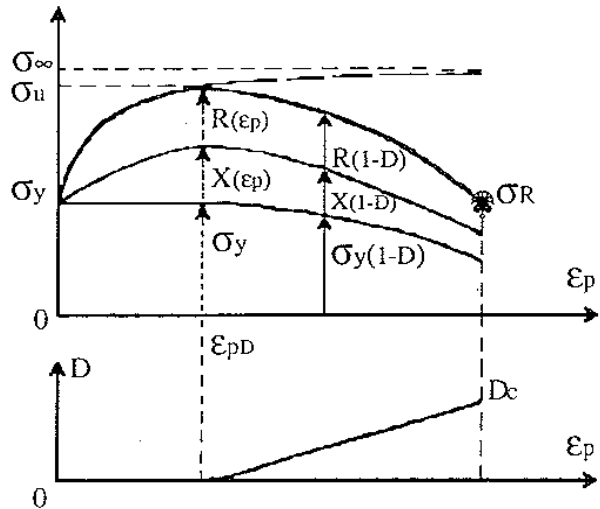

## 4.1 Algorithmic Implementation

In the *finite element method*, a given problem space is discretized into elements and further into nodes comprising these elements. The element shape is defined by shape functions of arbitrary polynomial degree, which allow for the interpolation of element shape between nodal locations. Because the element stiffness must be integrated over the element domain, the GAUSS integral allows for the necessary parameters to be determined at a finite number of GAUSS points. The global tangent stiffness matrix is then obtained by gather-scatter operations on the element stiffness matrices, using the elements' nodal connectivity as a map. For static problems the solution can be thought of as solving a generalized spring equation for the unknown forces and displacements.

While the exact solution to discretized linear problems is obtained in one single step, nonlinear problems are solved by additionally being time discretized, incurring incremental loading. By applying concepts of linearization to the problem-defining differential equation, its solution can be obtained via the NEWTON-RAPHSON method. The constitutive equations of the elements are evaluated at the GAUSS points at the time of element stiffness assembly. In the following, knowledge of the nonlinear finite element method is assumed, interested readers are therefore referred to BONET and WOOD (1997) [6] and BELYTSCHKO et al. (2000) [2].

In the case of material nonlinearity, these constitutive equations present an initial value problem, which is solved using the *backward* EULER *method* implicit pseudo-time integration scheme. For the case of the LEMAITRE damage model, the algorithm was developed by BENALLAL et al. (1988) [3] and later extended to *finite strains* by DE SOUZA NETO et al. (1994) [10]. The theory of the algorithm is summarized in the following. As the complete algorithmic implementation is rather extensive in nature, the reader is referred to the well-commented MATLAB code found in appendix B.2.

First, consider that the previous global NEWTON-RAPHSON iteration has computed a certain *elastic trial strain* (denoted by the index "tr"), which can be written as a part of an elastoplastic split so that

$$\boldsymbol{\varepsilon} = \boldsymbol{\varepsilon}_{\mathrm{tr}}^e + \boldsymbol{\varepsilon}_{\mathrm{tr}}^p \,. \tag{4.13}$$

This elastic trial strain can be seen as a prediction of the state of stress as a result of outdated internal variables. Therefore, this stage is also termed the *elastic predictor stage*, which aims to check the validity of the trial solution with respect to plastic admissibility. Also consider that the last iteration has yielded initial (denoted by the index 0) values of back-stress ($\boldsymbol{\beta}_0$), damage ($D_0$), and accumulated plastic strain ($\bar{\varepsilon}_0^p$) at the GAUSS points, let $R_0 = (1 - D_0)\, \bar{\varepsilon}_0^p$ and note that $\dot{R} = \dot{\lambda} = (1 - D)\, \dot{\bar{\varepsilon}}^p$ (cf. [3, 26]).

For each global NEWTON-RAPHSON iteration, the yield criterion is checked (see figure 2.3) at the GAUSS points:

$$\Phi_0 = \frac{q(\boldsymbol{\sigma}_d^{\mathrm{tr}} - \boldsymbol{\beta}_0)}{1 - D_0} - \kappa(R_0) - \sigma_{y0} \geq 0 \,. \tag{4.14}$$

If the yield criterion is not met, the elastic trial strain is accepted, the variables remain unaltered, and the damaged elastic tangent from equation 4.3 is returned. When yielding occurs, increments of the plastic multiplier $\Delta\lambda_c$ (index $c$ denoting the converged solution) are computed from the constitutive equations via a process called *return-mapping*. In geometric terms, the return-mapping procedure corresponds to an orthogonal deviatoric projection of the trial stresses onto the yield surface. It corrects the previously rejected elastic trial strain and is also termed the *plastic corrector stage* or *state variable update procedure*.

At the end of each iteration and after convergence of the return-mapping, the plastic strain ($\boldsymbol{\varepsilon}^p$) is corrected and the isotropic hardening internal variable ($R$) is updated so that

$$\boldsymbol{\varepsilon}^p = \boldsymbol{\varepsilon}_{\mathrm{tr}}^p + \Delta\boldsymbol{\varepsilon}_c^p = \boldsymbol{\varepsilon}_{\mathrm{tr}}^p + \Delta\lambda_c \boldsymbol{N}_c \,, \tag{4.15}$$

$$R = R_0 + (1 - D_c)\Delta\overline{\varepsilon}_c^p = R_0 + \Delta\lambda_c \,. \tag{4.16}$$

The other internal variables to update, the back-stress tensor ($\boldsymbol{\beta} = \boldsymbol{\beta}_c$) and the damage variable ($D = D_c$), are simply set to the values of the converged solution of the return-mapping procedure. The new true stress $\boldsymbol{\sigma}$ is computed from the corrected elastic strain $\boldsymbol{\varepsilon}^e = \boldsymbol{\varepsilon} - \boldsymbol{\varepsilon}^p$ and also set.

**A note on iterative indices:** The missing iterative indices above denote the variables' *current state* at the most recent known value of the internal variables. In this case, the most recent state is the converged solution. This notation applies to all iterative variables, including those contained in the following sections. Of course, missing indices in general may also denote a variable that stays untouched by the algorithm during each iteration, such as the total strain $\boldsymbol{\varepsilon}$. Constant variables are assumed to be known by the reader, as these follow from the physical interpretation of the problem and are not functions of the internal variables.

Finally, an updated tangent relation, termed the *consistent elastoplastic tangent modulus*, for the GAUSS point is obtained from the converged values of the internal variables and passed back to the computation of the new global tangent stiffness matrix. This step is crucial in ensuring quadratic convergence, as is explained in section 4.1.3.

This process is repeated until global convergence is achieved. As mentioned, the tangent stiffness matrix is recalculated at each global iteration, therefore this method represents the classical interpretation of the NEWTON-RAPHSON method, as opposed to the modified NEWTON-RAPHSON method, in which the tangent stiffness stays constant within each load step.

### 4.1.1 The Return-Mapping Procedure

As the return-mapping corresponds to a deviatoric projection of the trial stress onto the yield surface, one obvious criterion for the correct solution is that effective stress must be an element of the yield locus. The converged solution of the return-mapping then must fulfill $\Phi = 0$.

Additionally, the converged solution must satisfy the evolution equations for the internal variables. Thus, consider the pseudo-time discretized evolution equations of back-stress and damage

$$\Delta\boldsymbol{\beta} = \Delta\lambda(a\boldsymbol{N} - b\boldsymbol{\beta})\,, \tag{4.17}$$

$$\Delta D = \frac{\Delta\lambda}{1-D}\left(\frac{-Y}{r}\right)^s\,. \tag{4.18}$$

These equations can then be rewritten as

$$\boldsymbol{\beta} - \boldsymbol{\beta}_0 - \Delta\lambda(a\boldsymbol{N} - b\boldsymbol{\beta}) = \boldsymbol{0}\,, \tag{4.19}$$

$$D - D_0 - \frac{\Delta\lambda}{1-D}\left(\frac{-Y}{r}\right)^s = 0\,. \tag{4.20}$$

There are now three equations; to ensure completeness, recall that there are four independent variables contained in these equations, namely true stress ($\boldsymbol{\sigma}$), back-stress ($\boldsymbol{\beta}$), the increment of glide ($\Delta\lambda$), and the damage internal variable ($D$). In order to solve for the unknown independent variables, a fourth equation is required. This missing equation is the stress update relationship derived from equation 4.15 as follows:

$$\boldsymbol{\varepsilon}^p = \boldsymbol{\varepsilon} - \boldsymbol{\varepsilon}^e = \boldsymbol{\varepsilon} - \boldsymbol{\varepsilon}^e_{\text{tr}} + \Delta\lambda\boldsymbol{N} \tag{4.21}$$

$$\boldsymbol{0} = \boldsymbol{\varepsilon}^e - (\boldsymbol{\varepsilon}^e_{\text{tr}} - \Delta\lambda\boldsymbol{N}) \tag{4.22}$$

$$= \boldsymbol{\sigma} - (1-D)\mathbb{C}^e : (\boldsymbol{\varepsilon}^e_{\text{tr}} - \Delta\lambda\boldsymbol{N}) \tag{4.23}$$

The following coupled system of equations must now be solved for the internal variables $\boldsymbol{\gamma} = \{\boldsymbol{\sigma}, D, \Delta\lambda, \boldsymbol{\beta}\}$:

$$\begin{pmatrix} A^{\Phi} \\ \boldsymbol{A^{\sigma}} \\ \boldsymbol{A^{\beta}} \\ A^{D} \end{pmatrix} = \begin{pmatrix} \frac{q(\boldsymbol{\eta})}{1-D} - \kappa(R_0 + \Delta\lambda) - \sigma_{y0} \\ \boldsymbol{\sigma} - (1-D)\mathbb{C}^e : (\boldsymbol{\varepsilon}^e_{\text{tr}} - \Delta\lambda\boldsymbol{N}) \\ \boldsymbol{\beta} - \boldsymbol{\beta}_0 - \Delta\lambda(a\boldsymbol{N} - b\boldsymbol{\beta}) \\ D - D_0 - \frac{1}{1-D}\left(\frac{-Y}{r}\right)^s \Delta\lambda \end{pmatrix} = \begin{pmatrix} 0 \\ \boldsymbol{0} \\ \boldsymbol{0} \\ 0 \end{pmatrix}. \tag{4.24}$$

Recall that the flow vector and the damage strain energy release rate can also be written as

$$\boldsymbol{N}(\boldsymbol{\sigma}, \boldsymbol{\beta}, D) = \sqrt{\frac{3}{2}} \frac{\boldsymbol{\eta}}{(1-D)||\boldsymbol{\eta}||} = \frac{3}{2} \frac{\boldsymbol{\eta}}{(1-D)\, q(\boldsymbol{\eta})} \tag{4.25}$$

$$Y(\boldsymbol{\sigma}, D) = -\frac{1}{2(1-D)^2}\, \boldsymbol{\sigma} : \mathbb{C}^{e-1} : \boldsymbol{\sigma} \tag{4.26}$$

To solve this nonlinear problem, the system is linearized via TAYLOR-series expansion, neglecting higher order terms:

$$\boldsymbol{A}(\boldsymbol{\gamma}) \approx \boldsymbol{0} = \boldsymbol{A}(\boldsymbol{\gamma}_{-1}) + \frac{\partial \boldsymbol{A}}{\partial \boldsymbol{\gamma}} : \Delta\boldsymbol{\gamma} \tag{4.27}$$

$$= \left(\frac{\partial \boldsymbol{A}}{\partial \boldsymbol{\gamma}}\right)^{-1} : \boldsymbol{A}(\boldsymbol{\gamma}_{-1}) + (\boldsymbol{\gamma} - \boldsymbol{\gamma}_{-1}) \tag{4.28}$$

$$\boldsymbol{\gamma} = \boldsymbol{\gamma}_{-1} - \mathbb{J}^{-1} : \boldsymbol{A}(\boldsymbol{\gamma}_{-1}), \tag{4.29}$$

The increment in glide $\Delta\lambda$ is then iteratively computed by repeatedly solving equation 4.29, starting at $\boldsymbol{\gamma}_{-1} = \boldsymbol{\gamma}_0$ with an initial guess $\Delta\lambda_0 = 0$, and setting $\boldsymbol{\gamma}_{-1} = \boldsymbol{\gamma}$ at the beginning of each subsequent iteration. Here $\mathbb{J} = \frac{\partial \boldsymbol{A}}{\partial \boldsymbol{\gamma}}$ is the JACOBIAN matrix, $\boldsymbol{\gamma}$ is this step's solution, and $\boldsymbol{\gamma}_{-1}$ is the solution of the last iteration. All dependent variables are computed from $\boldsymbol{\gamma}_{-1}$ at the beginning of each iteration. This procedure is called the NEWTON-RAPHSON method and is followed until tolerance is met ($||\boldsymbol{A}(\boldsymbol{\gamma})|| \leq \text{tol} \approx 0$), when $\boldsymbol{\gamma}$ is regarded to be the converged solution. Finally, set $\boldsymbol{\gamma}_{\text{c}} = \boldsymbol{\gamma}$ and exit the return-mapping procedure.

### 4.1.2   Jacobian Matrix

To keep equations concise, define the integrity as $\omega = 1 - D$. The partial derivatives of $A^\Phi$ with respect to the internal variables $\boldsymbol{\sigma}, \Delta\lambda$, and $\boldsymbol{\beta}$ are

$$\boldsymbol{A}_{\boldsymbol{\sigma}}^{\Phi} = \frac{\partial A^\Phi}{\partial \boldsymbol{\sigma}} = \boldsymbol{N} = \frac{3}{2}\frac{\boldsymbol{\eta}}{\omega q}\,, \tag{4.30}$$

$$A_{\Delta\lambda}^{\Phi} = \frac{\partial A^\Phi}{\partial \Delta\lambda} = -\frac{\partial \kappa (R_0 + \Delta\lambda)}{\partial \Delta\lambda}\,, \tag{4.31}$$

$$\boldsymbol{A}_{\boldsymbol{\beta}}^{\Phi} = \frac{\partial A^\Phi}{\partial \boldsymbol{\beta}} = -\boldsymbol{N}\,. \tag{4.32}$$

Let the isotropic hardening thermodynamic force $\kappa$ be an exponential function of the form

$$\kappa(R_0 + \Delta\lambda) = R_\infty(1 - e^{-\gamma(R_0 + \Delta\lambda)})\,, \tag{4.33}$$

where $R_\infty$ is the asymptotic limit value of ultimate strength and $\gamma$ is an experimentally determined material parameter. Its derivative is then obtained to be

$$\frac{\partial \kappa(R_0 + \Delta\lambda)}{\partial \Delta\lambda} = \gamma R_\infty e^{-\gamma(R_0 + \Delta\lambda)}\,. \tag{4.34}$$

Differentiation with respect to $D$ requires more care to be taken. Recall that the true stress $\boldsymbol{\sigma}$ is given by equation 4.3 and is a function of $D$. Consider the derivative

$$\frac{\partial \boldsymbol{\sigma}}{\partial D} = -\mathbb{C}^e : \boldsymbol{\varepsilon}^e = -\boldsymbol{\sigma}_{\text{eff}} = -\frac{\boldsymbol{\sigma}}{\omega}\,. \tag{4.35}$$

Assuming $J_2$ to be a positive quantity, the derivatives of the relative stress and of the second invariant with respect to $D$ can be found:

$$\frac{\partial \boldsymbol{\eta}}{\partial D} = \frac{\partial \boldsymbol{\sigma}_d}{\partial D} = \frac{\partial \boldsymbol{\sigma}}{\partial D} - \frac{1}{3}\operatorname{tr}\left(\frac{\partial \boldsymbol{\sigma}}{\partial D}\right)\boldsymbol{1} = -\frac{\boldsymbol{\sigma}_d}{\omega}\,, \tag{4.36}$$

$$\frac{\partial J_2(\boldsymbol{\eta})}{\partial D} = \frac{1}{2}\frac{\partial}{\partial D}(\boldsymbol{\eta} : \boldsymbol{\eta}) = -\frac{1}{\omega}\,\boldsymbol{\sigma}_d : \boldsymbol{\eta}\,. \tag{4.37}$$

Note that the trace is a linear operator, therefore commuting with the derivative and that, due to the symmetry of the CAUCHY stress tensor, all contractions of two second order tensors in this section commute. The derivative of the VON MISES equivalent stress is then

given by

$$\frac{\partial q}{\partial D} = -\frac{3}{2} \frac{\boldsymbol{\sigma}_d : \boldsymbol{\eta}}{\omega q} .$$

(4.38)

By application of the product and chain rules, the derivative of $A^\Phi$ with respect to $D$ is

$$A_D^\Phi = \frac{\partial A^\Phi}{\partial D} = \frac{\partial}{\partial D} \frac{q}{\omega} = \frac{q}{\omega^2} - \frac{3}{2\omega^2 q} \boldsymbol{\sigma}_d : \boldsymbol{\eta} .$$

(4.39)

Finding the derivative of $\boldsymbol{A^\sigma}$ and $\boldsymbol{A^\beta}$ with respect to internal variables requires the derivatives of the flow vector $\boldsymbol{N}$. First, consider the following derivatives:

$$\frac{\partial \boldsymbol{\eta}}{\partial \boldsymbol{\sigma}} = \mathbb{1}^{\text{sym}} - \frac{1}{3} \mathbf{1} \otimes \mathbf{1} = \mathbb{1}^{\text{dev}} ,$$

(4.40)

$$\frac{\partial \boldsymbol{\eta}}{\partial \boldsymbol{\beta}} = -\mathbb{1}^{\text{sym}} ,$$

(4.41)

$$\frac{\partial q}{\partial \boldsymbol{\sigma}} = -\frac{\partial q}{\partial \boldsymbol{\beta}} = \frac{3}{2} \frac{\boldsymbol{\eta}}{q} ,$$

(4.42)

$$\frac{\partial q^{-1}}{\partial \boldsymbol{\sigma}} = -\frac{\partial q^{-1}}{\partial \boldsymbol{\beta}} = -\frac{3}{2} \frac{\boldsymbol{\eta}}{q^3} .$$

(4.43)

Then, for derivation with respect to $\boldsymbol{\sigma}$, the application of the product rule of differentiation to equation 4.25 yields

$$\frac{\partial \boldsymbol{N}}{\partial \boldsymbol{\sigma}} = \frac{3}{2\omega} \left( q^{-1} \frac{\partial \boldsymbol{\eta}}{\partial \boldsymbol{\sigma}} + \frac{\partial q^{-1}}{\partial \boldsymbol{\sigma}} \boldsymbol{\eta} \right)$$

(4.44)

$$= \frac{3}{2\omega q^3} \left( q^2 \, \mathbb{1}^{\text{dev}} - \frac{3}{2} \boldsymbol{\eta} \otimes \boldsymbol{\eta} \right) .$$

(4.45)

Obtaining the derivative of the flow vector with respect to $\boldsymbol{\beta}$ follows the same basic procedure, except for a change in sign and replacement of the deviatoric identity tensor $\mathbb{1}^{\text{dev}}$ with the symmetric identity tensor $\mathbb{1}^{\text{sym}}$. It follows that

$$\frac{\partial \boldsymbol{N}}{\partial \boldsymbol{\sigma}} = -\frac{3}{2\omega q^3} \left( q^2 \, \mathbb{1}^{\text{sym}} - \frac{3}{2} \boldsymbol{\eta} \otimes \boldsymbol{\eta} \right) .$$

(4.46)

Again, the derivative of the flow vector with respect to $D$ requires some work and is found

as follows:

$$\frac{\partial \boldsymbol{N}}{\partial D} = \frac{3}{2}\left(q^{-1}\frac{\partial}{\partial D}\left(\frac{\boldsymbol{\eta}}{\omega}\right) + \frac{\partial}{\partial D}\left(q^{-1}\right)\frac{\boldsymbol{\eta}}{\omega}\right) \tag{4.47}$$

$$= \frac{3}{2q}\frac{\boldsymbol{\eta} - \boldsymbol{\sigma}_d}{\omega^2} + \frac{9}{4q^3}\frac{(\boldsymbol{\sigma}_d : \boldsymbol{\eta})\boldsymbol{\eta}}{\omega^2} \tag{4.48}$$

$$= \frac{9}{4q^3}\frac{(\boldsymbol{\sigma}_d : \boldsymbol{\eta})\boldsymbol{\eta}}{\omega^2} - \frac{3}{2q}\frac{\boldsymbol{\beta}}{\omega^2}\,. \tag{4.49}$$

Consequently, the partial derivatives of $\boldsymbol{A^\sigma}$ with respect to the internal variables are

$$\mathbb{A}^{\boldsymbol{\sigma}}_{\boldsymbol{\sigma}} = \frac{\partial \boldsymbol{A^\sigma}}{\partial \boldsymbol{\sigma}} = \mathbb{1}^{\mathrm{sym}} + \omega\Delta\lambda\,\mathbb{C}^e : \frac{\partial \boldsymbol{N}}{\partial \boldsymbol{\sigma}}\,, \tag{4.50}$$

$$\boldsymbol{A}^{\boldsymbol{\sigma}}_D = \frac{\partial \boldsymbol{A^\sigma}}{\partial D} = \frac{\partial \boldsymbol{\sigma}}{\partial D} + \mathbb{C}^e : \left(\boldsymbol{\varepsilon}^e_{\mathrm{tr}} - \Delta\lambda\boldsymbol{N} + \omega\Delta\lambda\,\frac{\partial \boldsymbol{N}}{\partial D}\right)\,, \tag{4.51}$$

$$\boldsymbol{A}^{\boldsymbol{\sigma}}_{\Delta\lambda} = \frac{\partial \boldsymbol{A^\sigma}}{\partial \Delta\lambda} = \omega\mathbb{C}^e : \boldsymbol{N}\,, \tag{4.52}$$

$$\mathbb{A}^{\boldsymbol{\sigma}}_{\boldsymbol{\beta}} = \frac{\partial \boldsymbol{A^\sigma}}{\partial \boldsymbol{\beta}} = \omega\Delta\lambda\,\mathbb{C}^e : \frac{\partial \boldsymbol{N}}{\partial \boldsymbol{\beta}}\,. \tag{4.53}$$

The partial derivatives of $\boldsymbol{A^\beta}$ with respect to the internal variables are then given by the following equations:

$$\mathbb{A}^{\boldsymbol{\beta}}_{\boldsymbol{\sigma}} = \frac{\partial \boldsymbol{A^\beta}}{\partial \boldsymbol{\sigma}} = -a\Delta\lambda\frac{\partial \boldsymbol{N}}{\partial \boldsymbol{\sigma}}\,, \tag{4.54}$$

$$\boldsymbol{A}^{\boldsymbol{\beta}}_D = \frac{\partial \boldsymbol{A^\beta}}{\partial D} = -a\Delta\lambda\frac{\partial \boldsymbol{N}}{\partial D}\,, \tag{4.55}$$

$$\boldsymbol{A}^{\boldsymbol{\beta}}_{\Delta\lambda} = \frac{\partial \boldsymbol{A^\beta}}{\partial \Delta\lambda} = b\boldsymbol{\beta} - a\boldsymbol{N}\,, \tag{4.56}$$

$$\mathbb{A}^{\boldsymbol{\beta}}_{\boldsymbol{\beta}} = \frac{\partial \boldsymbol{A^\beta}}{\partial \boldsymbol{\beta}} = (1 + b\Delta\lambda)\mathbb{1}^{\mathrm{sym}} - a\Delta\lambda\frac{\partial \boldsymbol{N}}{\partial \boldsymbol{\beta}}\,. \tag{4.57}$$

In order to derive $A^D$ with respect to the internal variables, the corresponding derivatives of the damage strain energy release rate $Y$ must be known. Recall from equation 4.4 that $Y$ is not a function of damage, although it is expressed as such in equation 4.26. Therefore, only the derivative of $Y$ with respect to $\boldsymbol{\sigma}$ is required, while all other derivatives are null:

$$\frac{\partial Y}{\partial \boldsymbol{\sigma}} = -\frac{1}{\omega^2}\,\mathbb{C}^{e-1} : \boldsymbol{\sigma}\,. \tag{4.58}$$

Now, the derivatives of $A^D$ with respect to the internal variables are

$$\boldsymbol{A}_{\boldsymbol{\sigma}}^D = \frac{\partial A^D}{\partial \boldsymbol{\sigma}} = \frac{\Delta\lambda\, s}{r\omega}\, \frac{\partial Y}{\partial \boldsymbol{\sigma}} \left(\frac{-Y}{r}\right)^{s-1} , \tag{4.59}$$

$$A_D^D = \frac{\partial A^D}{\partial D} = 1 - \frac{\Delta\lambda}{\omega^2} \left(\frac{-Y}{r}\right)^s , \tag{4.60}$$

$$A_{\Delta\lambda}^D = \frac{\partial A^D}{\partial \Delta\lambda} = -\frac{1}{\omega} \left(\frac{-Y}{r}\right)^s , \tag{4.61}$$

$$\boldsymbol{A}_{\boldsymbol{\beta}}^D = \frac{\partial A^D}{\partial \boldsymbol{\beta}} = \boldsymbol{0} . \tag{4.62}$$

Finally, the above derivatives allow for the Jacobian to be written as

$$\mathbb{J} = \begin{pmatrix} \boldsymbol{A}_{\boldsymbol{\sigma}}^{\Phi} & A_D^{\Phi} & A_{\Delta\lambda}^{\Phi} & \boldsymbol{A}_{\boldsymbol{\beta}}^{\Phi} \\ \mathbb{A}_{\boldsymbol{\sigma}}^{\boldsymbol{\sigma}} & \boldsymbol{A}_D^{\boldsymbol{\sigma}} & \boldsymbol{A}_{\Delta\lambda}^{\boldsymbol{\sigma}} & \mathbb{A}_{\boldsymbol{\beta}}^{\boldsymbol{\sigma}} \\ \mathbb{A}_{\boldsymbol{\sigma}}^{\boldsymbol{\beta}} & \boldsymbol{A}_D^{\boldsymbol{\beta}} & \boldsymbol{A}_{\Delta\lambda}^{\boldsymbol{\beta}} & \mathbb{A}_{\boldsymbol{\beta}}^{\boldsymbol{\beta}} \\ \boldsymbol{A}_{\boldsymbol{\sigma}}^D & A_D^D & A_{\Delta\lambda}^D & \boldsymbol{A}_{\boldsymbol{\beta}}^D \end{pmatrix} . \tag{4.63}$$

### 4.1.3 The Consistent Elastoplastic Tangent Modulus

It has been shown by BENALLAL et al. (1988) [3] that the use of the consistent elastoplastic tangent modulus is critical in achieving near quadratic convergence at high values of damage, large load increments, and rapidly changing conditions. Such rapidly changing conditions introduce non-negligible second-order terms to the problem (cf. [12, 38]). In particular, the authors came to the conclusion that the use of the consistent tangent modulus improves global convergence by a factor of approximately two over the use of the general elastoplastic tangent operator or continuum tangent modulus. It is found that in the limit case, the increment in pseudotime ($\Delta t$) approaching zero, the general elastoplastic tangent coincides with the consistent elastoplastic tangent.

The derivation of the consistent elastoplastic tangent modulus from the Jacobian matrix as follows is implied in DE SOUZA NETO et al. (1994) [10]. It will be attempted to provide a step by step solution. The consistent elastoplastic tangent modulus can be obtained from taking the total derivative of $\boldsymbol{A}^{\sigma}$ with respect to the linear elastic strain tensor $\boldsymbol{\varepsilon}^e$

$$\frac{\mathrm{d}\boldsymbol{A}^{\sigma}}{\mathrm{d}\boldsymbol{\varepsilon}^e} = \frac{\partial \boldsymbol{A}^{\sigma}}{\partial \boldsymbol{\sigma}} : \frac{\mathrm{d}\boldsymbol{\sigma}}{\mathrm{d}\boldsymbol{\varepsilon}^e} + \frac{\partial \boldsymbol{A}^{\sigma}}{\partial D} \otimes \frac{\mathrm{d}D}{\mathrm{d}\boldsymbol{\varepsilon}^e} + \frac{\partial \boldsymbol{A}^{\sigma}}{\partial \Delta\lambda} \otimes \frac{\mathrm{d}\Delta\lambda}{\mathrm{d}\boldsymbol{\varepsilon}^e} + \frac{\partial \boldsymbol{A}^{\sigma}}{\partial \boldsymbol{\beta}} : \frac{\mathrm{d}\boldsymbol{\beta}}{\mathrm{d}\boldsymbol{\varepsilon}^e} . \tag{4.64}$$

The only total derivative on the right hand side of the equation that is not null is $\mathrm{d}\boldsymbol{\sigma}/\mathrm{d}\boldsymbol{\varepsilon}^e$.

Multiplying the equation by $\mathrm{d}\boldsymbol{\varepsilon}^e$, the infinitesimal differential

$$\mathrm{d}\boldsymbol{A^\sigma} = \frac{\partial \boldsymbol{A^\sigma}}{\partial \boldsymbol{\sigma}} : \mathrm{d}\boldsymbol{\sigma} \tag{4.65}$$

is found. Recalling equations 4.15 and 4.50, the above infinitesimal can also be written as

$$\mathrm{d}\boldsymbol{A^\sigma} = \frac{\partial \boldsymbol{\sigma}}{\partial \boldsymbol{\sigma}} : \mathrm{d}\boldsymbol{\sigma} + \omega \mathbb{C}^e : \left( \Delta\lambda \frac{\partial \boldsymbol{N}}{\partial \boldsymbol{\sigma}} : \mathrm{d}\boldsymbol{\sigma} \right) \tag{4.66}$$

$$= \mathrm{d}\boldsymbol{\sigma} + \omega \mathbb{C}^e : \mathrm{d}\Delta\boldsymbol{\varepsilon}^p \,. \tag{4.67}$$

From the elastoplastic split of the strain tensor and equation 4.15, it is known that

$$\mathrm{d}\Delta\boldsymbol{\varepsilon}^p = -\mathrm{d}\Delta\boldsymbol{\varepsilon}^e = \mathrm{d}\boldsymbol{\varepsilon}^e_{\mathrm{tr}} - \mathrm{d}\boldsymbol{\varepsilon}^e \,. \tag{4.68}$$

The damaged elastic law implies $\mathrm{d}\boldsymbol{\sigma} = \omega \mathbb{C}^e : \mathrm{d}\boldsymbol{\varepsilon}^e$, therefore

$$\mathrm{d}\boldsymbol{A^\sigma} = \omega \mathbb{C}^e : (\mathrm{d}\boldsymbol{\varepsilon}^e - \mathrm{d}\Delta\boldsymbol{\varepsilon}^e) \tag{4.69}$$

$$= \omega \mathbb{C}^e : \mathrm{d}\boldsymbol{\varepsilon}^e_{\mathrm{tr}} \,. \tag{4.70}$$

With the internal variables $\boldsymbol{\gamma}$ and the return-mapping system of equations $\boldsymbol{A}$, the infinitesimal of $\boldsymbol{A^\sigma}$ is

$$\mathrm{d}\boldsymbol{A^\sigma} = \mathrm{d}\boldsymbol{A}_2 = \mathbb{J}_{2j}\,\mathrm{d}\boldsymbol{\gamma}_j \tag{4.71}$$

$$= \mathbb{J}_{21} : \mathrm{d}\boldsymbol{\sigma} \,. \tag{4.72}$$

The consistent elastoplastic tangent modulus (cf. [10]) is given by

$$\mathbb{D}^{ep}_c = \frac{\mathrm{d}\boldsymbol{\sigma}}{\mathrm{d}\boldsymbol{\varepsilon}^e_{\mathrm{tr}}} \,. \tag{4.73}$$

Finally, combining equations 4.70 and 4.72 yields

$$\mathbb{J}_{21} : \mathrm{d}\boldsymbol{\sigma} = \omega \mathbb{C}^e : \mathrm{d}\boldsymbol{\varepsilon}^e_{\mathrm{tr}} \tag{4.74}$$

$$\mathbb{D}^{ep}_c = \frac{\mathrm{d}\boldsymbol{\sigma}}{\mathrm{d}\boldsymbol{\varepsilon}^e_{\mathrm{tr}}} = \omega \mathbb{J}^{-1}_{12} : \mathbb{C}^e \,. \tag{4.75}$$

Note that, due to the coupling of damage, $\mathbb{D}^{ep}_c$ is only minor symmetric (cf [11,12]). A closed form solution for the consistent elastoplastic modulus that does not require matrix inversion was derived by DOGHRI (1995) [12]. For LEMAITRE's simplified damage model, a closed form solution also exists (cf. [11]). As it includes fairly lengthy coefficients, these are omitted at this point but can be found in the source code of the model (see appendix B.1) and in the reference work. The final relation for the simplified model is

$$
\mathbb{D}_c^{ep} = a\,\mathbb{1}^{\text{dev}} + b\,\frac{\boldsymbol{\sigma}_d}{||\boldsymbol{\sigma}_d||} \otimes \frac{\boldsymbol{\sigma}_d}{||\boldsymbol{\sigma}_d||} \;+\; c\,\frac{\boldsymbol{\sigma}_d}{||\boldsymbol{\sigma}_d||} \otimes \mathbf{1} + d\,\mathbf{1} \otimes \frac{\boldsymbol{\sigma}_d}{||\boldsymbol{\sigma}_d||} + e\,\mathbf{1} \otimes \mathbf{1}\,, \tag{4.76}
$$

where $a, b, c, d$, and $e$ are functions of $D$, $q_{\text{tr}}$, $p_{\text{tr}}$, $H = \mathrm{d}\sigma_{y0}/\mathrm{d}R$, $\kappa(R)$, and $\sigma_{y0}$.

## 4.2  Lemaitre's Simplified Damage Model

LEMAITRE's damage model can be simplified by the absence of kinematic hardening. This allows for the return-mapping system of equations to be reduced to a single scalar equation (cf. [11]). Its derivation is a rather lengthy procedure, thus only the most important relations are summarized in the following.

The absence of kinematic hardening reduces $\Phi$ to

$$
\Phi = \frac{q(\boldsymbol{\sigma}_d)}{\omega} - \kappa(R_0 + \Delta\lambda) - \sigma_{y0}\,. \tag{4.77}
$$

According to equation 4.15, the update relation for the deviatoric elastic strain is

$$
\boldsymbol{\varepsilon}_d^e = \boldsymbol{\varepsilon}_{d,\text{tr}}^e - \Delta\lambda\,\boldsymbol{N}(\boldsymbol{\sigma}_d, D)\,. \tag{4.78}
$$

Recall that volumetric strain remains unchanged due to isochoric yielding and therefore is simply $\boldsymbol{\varepsilon}_h^e = \boldsymbol{\varepsilon}_{h,\text{tr}}^e = \mathrm{tr}(\boldsymbol{\varepsilon}_{\text{tr}}^e)$. Consequently, the hydrostatic stress is $p = p_{\text{tr}} = \omega K\,\mathrm{tr}(\boldsymbol{\varepsilon}_{\text{tr}}^e)$. The update relations for the hardening internal variable $R$ and damage are given by equations 4.16 and 4.20, respectively.

Application of the elastic law to equation 4.78 yields

$$
\omega\,2G\,\boldsymbol{\varepsilon}_d^e = \omega\,2G\,\boldsymbol{\varepsilon}_{d,\text{tr}}^e - \omega\,2G\,\Delta\lambda\,\boldsymbol{N}(\boldsymbol{\sigma}_d, D) \tag{4.79}
$$

$$
\boldsymbol{\sigma}_d = \omega\,\boldsymbol{\sigma}_{\text{eff},d}^{\text{tr}} - 3G\,\Delta\lambda\,\frac{\boldsymbol{\sigma}_d}{q(\boldsymbol{\sigma}_d)}\,. \tag{4.80}
$$

Grouping terms in the above equation shows that $\boldsymbol{\sigma}_d \propto \boldsymbol{\sigma}_{\text{eff},d}^{\text{tr}}$, thus, as $\boldsymbol{\sigma}_d$ only occurs normalized in the flow vector, it can be replaced by $\boldsymbol{\sigma}_{\text{eff},d}^{\text{tr}}$. The VON MISES stress then must

be

$$q(\boldsymbol{\sigma}_d) = \sqrt{\frac{3}{2}\left(\omega - 3G\frac{\Delta\lambda}{q(\boldsymbol{\sigma}_{\text{eff},d}^{\text{tr}})}\right)^2 \boldsymbol{\sigma}_{\text{eff},d}^{\text{tr}} : \boldsymbol{\sigma}_{\text{eff},d}^{\text{tr}}} \qquad (4.81)$$

$$= \omega q(\boldsymbol{\sigma}_{\text{eff},d}^{\text{tr}}) - 3G\Delta\lambda\,. \qquad (4.82)$$

With the above equation the yield function is

$$\Phi = 0 = q(\boldsymbol{\sigma}_{\text{eff},d}^{\text{tr}}) - 3G\frac{\Delta\lambda}{\omega} - \kappa(R_0 + \Delta\lambda) - \sigma_{y0}\,. \qquad (4.83)$$

Solving for the material's integrity,

$$\omega(\Delta\lambda) = \frac{3G\,\Delta\lambda}{q(\boldsymbol{\sigma}_{\text{eff},d}^{\text{tr}}) - \kappa(R_0 + \Delta\lambda) - \sigma_{y0}} \qquad (4.84)$$

as a function of the increment in glide ($\Delta\lambda$) alone is obtained. By some algebra (cf. [11]) and the replacement of the conventional VON MISES stress with equation 4.82, the damage strain energy release rate from equation 4.26 can also be written as

$$Y(\Delta\lambda) = -\frac{(\kappa(R_0 + \Delta\lambda) + \sigma_{y0})^2}{6G} - \frac{(p_{\text{eff}}^{\text{tr}})^2}{2K}\,. \qquad (4.85)$$

Combining the latter equation with equation 4.84 and the damage evolution law from equation 4.20 (cf. [11]), the update equation constituting the single equation return-mapping is found to be

$$A^D(\Delta\lambda) = 0 = \omega(\Delta\lambda) - \omega_0 + \frac{\Delta\lambda}{\omega(\Delta\lambda)}\left(\frac{-Y(\Delta\lambda)}{r}\right)^s\,. \qquad (4.86)$$

Its derivative with respect to the increment in glide is

$$A_{\Delta\lambda}^D = \frac{\mathrm{d}A^D}{\mathrm{d}\Delta\lambda} = y + \Delta\lambda\frac{3G}{(q(\boldsymbol{\sigma}_{\text{eff},d}^{\text{tr}}) - \kappa - \sigma_{y0})^2}\kappa_{\Delta\lambda} \qquad (4.87)$$

$$- \frac{\kappa_{\Delta\lambda}}{3G}\left(\frac{-Y}{r}\right)^s - \frac{sY_{\Delta\lambda}}{ry}\left(\frac{-Y}{r}\right)^{s-1}\,, \qquad (4.88)$$

$$Y_{\Delta\lambda} = \frac{\mathrm{d}Y}{\mathrm{d}\Delta\lambda} = -\frac{(\kappa + \sigma_{y0})\kappa_{\Delta\lambda}}{3G}\,, \qquad (4.89)$$

$$y = \frac{3G}{q(\boldsymbol{\sigma}_{\text{eff},d}^{\text{tr}}) - \kappa - \sigma_{y0}}\,. \qquad (4.90)$$

The solution for $\Delta\lambda$ in the update equation is obtained via the NEWTON-RAPHSON method as outlined in section 4.1.1. Again, the solution to the dependent variables $\omega$, $\kappa$, and $Y$ is obtained at the beginning of the current iteration from the result of the last iteration. An initial guess

$$\Delta\lambda_0 = \frac{\omega_0(q(\boldsymbol{\sigma}_{\text{eff},d}^{\text{tr}}) - \kappa - \sigma_{y0})}{3G} \tag{4.91}$$

is employed in order to improve convergence over the use of $\Delta\lambda_0 = 0$ (cf. [11]).

After the procedure has converged, the VON MISES stress, true stress, and plastic strain are updated as follows from the equations above so that

$$q(\boldsymbol{\sigma}_d) = \omega_c\,\kappa\,(R_0 + \Delta\lambda_c) \tag{4.92}$$

$$p = \omega_c\,p_{\text{eff}}^{\text{tr}}\,, \tag{4.93}$$

$$\boldsymbol{\sigma}_d = 2G\,\frac{q(\boldsymbol{\sigma}_d)}{q(\boldsymbol{\sigma}_{\text{eff},d}^{\text{tr}})}\,\boldsymbol{\varepsilon}_{d,\text{tr}}^e\,, \tag{4.94}$$

$$\boldsymbol{\varepsilon}^p = \boldsymbol{\varepsilon} - \frac{\boldsymbol{\sigma}_d}{\omega_c\,2G} - \frac{1}{3}\,\frac{p_{\text{eff}}^{\text{tr}}}{K}\,\mathbf{1}\,. \tag{4.95}$$

Finally, the consistent elastoplastic tangent modulus is returned as outlined in sections 4.1 and 4.1.3.

This return-mapping algorithm and the MATLAB implementation of LEMAITRE's simplified damage model (see appendix B.1) are based on the procedure outlined in the source code of the finite element program HYPLAS, developed by DE SOUZA NETO et al. and included in their book published in 2008 [11].

## 4.3 Issues Arising From Voigt Notation

Tensors of order $k$ and dimension $n$ contain $n^k$ elements, while symmetric second order tensors possess $(n^2 + n)/2$ unique elements. Fourth order tensors constructed from the dyadic product of two symmetric second order tensors have $((n^2 + n)/2)^2$ unique elements and are classified as *minor symmetric* ($\mathbb{A}_{ijkl} = \mathbb{A}_{jilk}$). A minor symmetric three-dimensional tensor of fourth order therefore contains 36 unique elements. In special cases a fourth order tensor exhibits *major symmetry* so that $\mathbb{A}_{ijkl} = \mathbb{A}_{klij}$. A fourth order tensor of dimension three then has 45 unique elements. Fourth order tensors possessing both minor and major symmetries are also termed *supersymmetric* and, for the case of three dimensions, contain 21 unique elements.

Symmetric tensors allow for the reduction of the tensor to its unique elements. Second order tensors therefore can be expressed as vector-like structures, while fourth order tensors reduce to matrix-like structures. Such an expression of symmetric tensors is called VOIGT notation. These vector- or matrix-like structures do not transform as vectors or matrices under a change of basis, nor do simple vector operations apply, as the original tensor is defined in a higher-order vector space. From such a definition, a series of problems can arise in implementing standard tensor expressions and derivatives in VOIGT notation.

Specifically, the expression of tensor contractions in VOIGT notation involves the need for the introduction of scaling factors for the off-diagonal elements to the calculation. This need arises as the original tensor contains the off-diagonal elements twice. For example, the double tensor contraction of a symmetric tensor $\boldsymbol{A} \in \mathbb{R}^2$ with itself is $\boldsymbol{A}_{ij}\boldsymbol{A}_{ij} = a_{11}^2 + a_{22}^2 + 2a_{12}^2$. If $\boldsymbol{A}$ were reduced to a true vector $\vec{a}$, the single contraction or scalar product $\vec{a} \cdot \vec{a}$ would yield $\vec{a}_\alpha \vec{a}_\alpha = a_{11}^2 + a_{22}^2 + a_{12}^2$.

For both operations to be equal, two approaches are possible. The first is the use of MANDEL notation, involving a scaling of the off-diagonal elements by the factor $\sqrt{2}$. This alters the original values of the off-diagonal elements and can be difficult to keep track of in complex calculations. The second approach is to use the HADAMARD (or entrywise) product ($(\vec{a} \circ \vec{b})_\alpha = \vec{a}_{,\alpha} \cdot \vec{b}_{,\alpha}$; the comma implying there being no summation over the index) and to define a scaling vector $\vec{s} = [1\ 1\ 2]^{\mathrm{T}}$ so that $\boldsymbol{A}_{ij}\boldsymbol{A}_{ij} = \sum\limits_{\alpha}((\vec{a} \circ \vec{a}) \circ \vec{s})_\alpha$.

To satisfy the relation given by equation 2.4 so that $2\rho\psi^e = \boldsymbol{\varepsilon}^V \cdot \boldsymbol{\sigma}^V$, the representation of the linear strain tensor in VOIGT notation introduces a scaling factor of two to the off-diagonal elements. The scaled shear components of the strain tensor in VOIGT notation are termed *engineering strain* so, for example, $\gamma_{12} = 2\varepsilon_{12}$. The CAUCHY stress tensor is written in VOIGT notation without scaling factors applied.

The same problem arises when attempting to write a double tensor contraction of a fourth and second order tensor as a vector operation. This is of particular importance for the calculation of HOOKE's law $\boldsymbol{\sigma} = \mathbb{C}^e : \boldsymbol{\varepsilon}$. With the definition of the linear strain tensor in VOIGT notation including engineering strain, HOOKE's law can be expressed as $\boldsymbol{\sigma}^V = \mathbb{C}^V \cdot \boldsymbol{\varepsilon}^V$, avoiding the introduction of scaling factors to $\mathbb{C}^V$. In general, the relation is $\boldsymbol{C}_\alpha^V = (\mathbb{B}_{\alpha\beta}^V \boldsymbol{A}_\beta^V)_{,\alpha}\ \vec{s}_{,\alpha} = (B \cdot \vec{a}) \circ \vec{s}$.

The dyadic product $\mathbb{D} = \boldsymbol{A} \otimes \boldsymbol{B}$ in VOIGT notation does not require the use of scaling factors. Assuming that $\boldsymbol{A}$ and $\boldsymbol{B}$ are written in VOIGT notation without scaling present, the latter relation can also be written as $\mathbb{D}_{\alpha\beta}^V = \boldsymbol{A}_\alpha^V \boldsymbol{B}_\beta^V = \vec{a} \cdot \vec{b}^{\,\mathrm{T}}$.

Another issue that deserves particular attention in this context is the writing of tensor derivatives in VOIGT notation. Consider the partial derivative of the CAUCHY stress tensor in VOIGT notation with respect to its tensorial notation counterpart $\partial\boldsymbol{\sigma}^V_\alpha/\partial\boldsymbol{\sigma}_{kl}$. In tensorial notation, this derivative would evaluate to symmetric unity $\mathbb{1}^{\text{sym}}$. In general, fourth order symmetric identity in VOIGT notation can be written as

$$\mathbb{1}^{\text{sym},V} = \begin{pmatrix} 1 & 0 & 0 \\ 0 & 1 & 0 \\ 0 & 0 & \frac{1}{2} \end{pmatrix}, \tag{4.96}$$

where the value of $1/2$ corresponds to the position of the off-diagonal components. For simplicity, only two diagonal components and one off-diagonal component are shown. Yet, $\partial\boldsymbol{\sigma}^V/\partial\boldsymbol{\sigma} = \mathbb{1}^{\text{sym},V}$ is not the correct solution as is explained in the following.

If the symmetric identity tensor is a result of a tensor derivative such as the one considered above, *the off-diagonal components of* $\mathbb{1}^{\text{sym},V}$ *must be multiplied by a factor of two.* This follows from differentiation of a tensor in VOIGT notation with respect to a symmetric tensor as, in this case, the derivatives with respect to the off-diagonal components occur twice so $\partial\boldsymbol{\sigma}^V_{12}/\partial\boldsymbol{\sigma}_{12} = \partial\boldsymbol{\sigma}^V_{12}/\partial\boldsymbol{\sigma}_{21} = 1/2$, whereas $\partial\boldsymbol{\sigma}_{12}/\partial\boldsymbol{\sigma}_{12} = 1$. This equally applies to derivatives in tensorial notation resulting in the fourth order deviatoric identity tensor $\mathbb{1}^{\text{dev}}$. For the purpose of distinguishing the identity tensors resulting from differentiation, in the following they are marked as $\hat{\mathbb{1}}^{\text{sym},V}$ and $\hat{\mathbb{1}}^{\text{dev},V}$. Note that, while $\hat{\mathbb{1}}^{\text{dev},V}_{\alpha\beta}\text{dev}[\boldsymbol{A}]^V_\beta = \text{dev}[\boldsymbol{A}]^V_\alpha$ preserves the deviator, $\mathbb{1}^{\text{dev},V}_{\alpha\beta}\text{dev}[\boldsymbol{A}]^V_\beta \neq \text{dev}[\boldsymbol{A}]^V_\alpha$ does not. The analogue is true for the same operation on symmetric unity.

Similarly, the tensor obtained from derivation of a scalar with respect to a tensor and written in VOIGT notation incurs multiplication of the off-diagonal components by a factor of two. As an example, consider the partial derivative of the second invariant $J_2(\boldsymbol{\eta}^V)$ of relative stress in VOIGT notation with respect to the CAUCHY stress tensor $\boldsymbol{\sigma}_{ij}$. The second invariant in VOIGT notation is given by

$$J_2(\boldsymbol{\eta}^V) = \frac{1}{2}\,\boldsymbol{\eta}^V \circ \boldsymbol{\eta}^V \circ \vec{s}. \tag{4.97}$$

The partial derivative then is

$$\left(\frac{\partial J_2(\boldsymbol{\eta}^V)}{\partial\boldsymbol{\sigma}_{ij}}\right)_\alpha = \left(\hat{\mathbb{1}}^{\text{dev},V}_{\alpha\beta}\boldsymbol{\eta}^V_\beta\right)_{,\alpha}\,\vec{s}_{,\alpha}. \tag{4.98}$$

As the product $\hat{\mathbb{1}}^{\text{dev},V}_{\alpha\beta}\boldsymbol{\eta}^V_\beta$ preserves the same relation as $\mathbb{1}^{\text{dev}} : \boldsymbol{\eta} = \boldsymbol{\eta}$, the result is

$$\frac{\partial J_2(\boldsymbol{\eta}^V)}{\partial \boldsymbol{\sigma}} = \boldsymbol{\eta}^V \circ \vec{s}. \tag{4.99}$$

In contrast to the issues outlined above, the use of Voigt notation in the computational implementation of continuum mechanics problems has distinct advantages over the use of the Einstein summation convention. The reduction to one or two dimensional arrays leads to much smaller memory overhead; in addition, tensor operations in Voigt notation greatly reduce the number of indexing operations. For example, the double contraction of a second order and a fourth order tensor $((\mathbb{A} : \boldsymbol{B})_{ij} = \mathbb{A}_{ijkl}\boldsymbol{B}_{kl})$, using the Einstein summation convention, involves two independent indices and two dummy indices, requiring four nested loops and $n^4$ operations for tensors of dimension $n$. In Voigt notation, assuming the appropriate scaling factors are present, the same operation can be expressed as the multiplication of a matrix and a vector $((\mathbb{A} : \boldsymbol{B})^V_\alpha = \mathbb{A}^V_{\alpha\beta}\boldsymbol{B}^V_\beta)$, therefore only requiring two nested loops and $((n^2 + n)/2)^2$ operations.

## 4.4   Verification of the Models

In order to verify the correct implementation of the full and simplified versions of Lemaitre's damage model, calculations with the finite element method were carried out. Results from the simplified version of Lemaitre's damage model are presented in de Souza Neto et al. (2008) [11]. In the latter work, calculations on a cylindrical notched specimen meshed with eight-node axisymmetric quadrilateral elements were performed.

To provide a basis for comparison, the same geometry was created in and meshed with the general-purpose commercial finite element program ANSYS. For appropriate mesh sizing control and the high probability of achieving mesh integrity, a mapped mesh was chosen, in contrast to the free mesh employed in the reference work. The geometry of a quarter of the cylindrical notched specimen as well as a coarse and a fine mesh are shown in figure A.1. Calculations were carried out within an existing proprietary finite element framework written in MATLAB and developed by the Institute of Mechanics and Computational Mechanics at the University of Hannover, Germany. In order to model axisymmetry, appropriate eight-node quadrilateral elements and the corresponding circumferential element stiffness integration were implemented. The open-source finite element software package Gmsh [16] was used for parts of the post-processing stage of the analysis.

Boundary conditions were chosen according to the rules of symmetry: the nodes on the left edge were constrained in horizontal direction, while the nodes on the bottom edge were constrained in vertical direction. An edge displacement value of one in vertical direction was imposed on the nodes at the top edge of the model to allow for displacement control within the calculation. The material parameters used are listed in table 4.1. The parameters, except those for the ARMSTRONG-FREDERICK kinematic hardening law ($a$ and $b$), were calibrated by BENALLAL et al. (1987) [4] for AISI 1010 (DIN CK10) rolled low carbon steel (cf. [11]). The parameters $a$ and $b$ were obtained from the work of BENALLAL and coworkers (1988) [3]. The boundary conditions, displacements, and the material parameters required for the simplified model correspond to those of the reference work.

Table 4.1: Material parameters for LEMAITRE's damage model.

| | |
|---|---|
| $E$ | 210 GPa |
| $\nu$ | 0.3 |
| $\sigma_{y0}$ | 620 MPa |
| $R_\infty$ | 3300 MPa |
| $\gamma$ | 0.4 |
| $a$ | 2500 MPa |
| $b$ | 20 |
| $r$ | 3.5 MPa |
| $s$ | 1 |

The tensile loading computations were performed by imposing a maximum upper edge displacement of $u = 0.576$ mm for the simplified model and $u = 0.656$ mm for the model including kinematic hardening. Both calculations were subject to the same load step increments below $u = 0.57$ mm; a total of 60 increments were employed to attain this displacement. For the simplified model, an additional 20 increments were chosen to attain a maximum displacement of $u = 0.576$ mm. The model including kinematic hardening was subject to another 50 increments above $u = 0.57$ mm to attain a maximum displacement of $u = 0.656$ mm. The maximum edge displacement for the simplified model corresponds to that of the reference work.

Maximum GAUSS point damage (element 1, GAUSS point 9 - location closest to the center of the specimen) is plotted over upper edge displacement in figure A.3. As the reference work contains similar data from the center of the model, this was compared to the data obtained from the tensile loading computation carried out with the implementation of LEMAITRE's simplified model. It can be seen in figure A.3 that results show fairly good agreement for the majority of the displacement range, although asymptotic behavior of damage appears to be protracted in computations with this work's implementation.

Initially, it must be recognized that most damage models are known to be mesh-dependent. This is a result of the use of the principle of strain equivalence introducing a softening contribution to the material's yield stress and therefore requiring a length scale (cf. [7]). As the mesh used in the reference work does not correspond to the meshes generated by the author, such mesh effects are expected to be significant.

It should be kept in mind that the "center of the model" is only weakly defined and could be a location of an edge node in the reference work's implementation. As the considered GAUSS point is not exactly in the center of the specimen, the minor deviations observed are possibly a result of data originating from different locations within the model. It should also be noted that the calculation from which the data was obtained was carried out on a mesh comprising a significantly lesser amount of elements. This is expected to primarily contribute to deviations as a result of mesh-size-dependence as well as GAUSS point distance to the center of the specimen being relatively large.

In this context, the nodal damage plots of LEMAITRE's simplified model (see figure A.5) should be considered. It must be acknowledged that damage dependence on stress triaxiality ratio $p/q$ is modeled correctly by the implementation. As expected from experimental observations (cf. [18]), damage localization and therefore eventual crack initiation is predicted at the center of the cylindrical specimen. This is supported by levels of stress triaxiality ratio shown in figure A.7. It can be observed that, as damage progresses, triaxiality ratio increases. As a result of this increase in triaxiality, the material experiences loss of ductility and will finally fail when accumulated plastic strain reaches a critical level (cf. [18]).

In comparison to the reference work (see figure A.6), the shapes of the isolines appear identical. Minor deviations in the maximum nodal damage values are observed for the plots $u = [0.051; 0.076; 0.246]\,\text{mm}$, while significant deviation can be seen in the final plot at $u = 0.576\,\text{mm}$. Mesh-dependence is assumed by the author to also be damage dependent, resulting in higher deviations from results obtained from a different mesh as damage progresses. This could be a result of the softening effect introduced by the concept of effective stress, which is dependent on damage.

This softening effect can be observed clearly in plots showing the true stress plotted over displacement (see figure A.4 for the simplified and A.8 for the model including kinematic hardening). It can be seen that damage equally reduces the isotropic hardening thermody-namic force $\kappa$, the back-stress $\boldsymbol{\beta}$, as well as the initial yield stress $\sigma_{y0}$, as is expected from the definition of the model using LEMAITRE's hypothesis of strain equivalence.

Differences in comparison with the stress-strain-curves from figure 4.1 arise by the absence of kinematic hardening in the simplified model, as well as the implementations' disregard for the damage threshold. Within the implementations developed by the author as well as the one contained in the reference work, damage nucleation occurs at the strain required for yielding. Such an implementation therefore does not model the physical assumption that damage nucleation only occurs above a certain damage threshold, measured in accumulated plastic strain. For details on its implementation, the reader is referred to the reference work.

Ultimately, results for both the simplified and full implementations of LEMAITRE's damage model can be considered to be in satisfactory agreement with the expected stress-strain behavior, considering the limitations discussed above. However, it is concluded that further investigations and comparison with analytical solutions must be undertaken to sufficiently verify correct behavior.

A comparison of damage evolution in the calculations carried out with the simplified and full implementations is shown in figure A.2. As expected, damage evolves less rapidly in the full implementation's computation as a result of increased hardening attributed to the presence of back-stress. This can also be observed by comparing the nodal damage plots of both models (see figures A.5 and A.10). The nodal damage plot of the kinematic hardening model (figure A.10) reveals the spatial effect of back-stress on damage distribution. From the elongated elliptical geometry of the isolines, in contrast to those of the simplified model, it can clearly be seen that the movement of damage localization is slowed by back-stress evolution.

This behavior is expected from the microstructural causes of the BAUSCHINGER effect as discussed in section 2.3.2. The observation of slowed movement of damage localization is supported by the plots of nodal back-stress, which can be seen in figure A.12. To further study the effect of back-stress, cyclic loading computations were carried out. Displacement control was used to achieve two full cycles of $u = 0.076\,\text{mm}$ at a displacement ratio of $R_u = -1$. It can be seen from figure A.9 that loading in compression correctly models the decrease in yield stress due to the aide of released back-stress. Isotropic and kinematic hardening are visible in the translation and dilation of the yield surface.

## 4.5   Convergence

The MATLAB implementation of the simplified model shows quadratic levels of global convergence throughout the calculation. The error in the prediction of internal forces is in the range of $10^{-8}$ and $10^{-4}$, as can be seen in figure 4.2. The maximum number of iterations needed for achieving the convergence requirement (error in displacement prediction less than $10^{-9}$) was six, although convergence within five iterations was usually obtained (see figure 4.4). Convergence of the return-mapping was regularly achieved in four to five steps. There was no dynamic stepping employed. These observations appear to be an indication of correct implementation.

Figure 4.2: Convergence of the MATLAB implementation of LEMAITRE's simplified damage model. Calculation of model with 121 nodes, 32 elements.

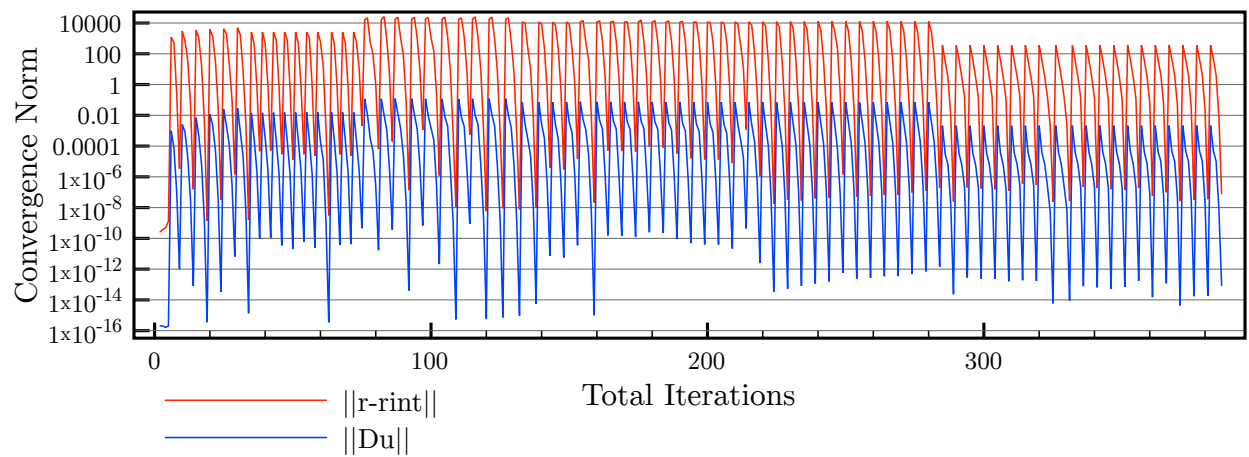

It can be seen that global convergence of the model including kinematic hardening deteriorates with damage. The error in internals was in the range of $10^{-5}$ and $10^{-4}$ (see figure 4.3), while the maximum number of iterations needed for convergence was 17 at the very end of the computation (see figure 4.4). Usually, convergence was achieved within eight iterations. The return-mapping algorithm also exhibited questionable levels of convergence. It is observed that the number of steps needed to achieve convergence lies between four and twelve, depending on damage. There was no convergence control in place, nor were line-search procedures implemented. From the plot of the convergence norms (figure 4.3), it can be seen that as damage localization begins to move towards the center of the specimen, larger initial errors are encountered. Possibly, this could be a problem induced by linearization as has been discussed in section 4.1.3.

Deterioration of convergence was observed to be highly significant at states of damage above $D = 0.4$. Although it is known that convergence deteriorates with damage (cf. [11]), BENALLAL et al. (2008) [3] achieved convergence within ten steps by use of the consistent elastoplastic tangent. It must be noted that a different geometry was used and that loading was cyclic and of bending nature, which is therefore not representative for the currently considered specimen. As a result of not having modeled a geometry as found in BENALLAL et al. (2008) [3], it is difficult to analyze the reasons for this non-quadratic convergence, in consequence, it is possible that errors were made in the implementation.

Figure 4.3: Convergence of the MATLAB implementation of LEMAITRE's damage model including kinematic hardening. Calculation of model with 121 nodes, 32 elements.

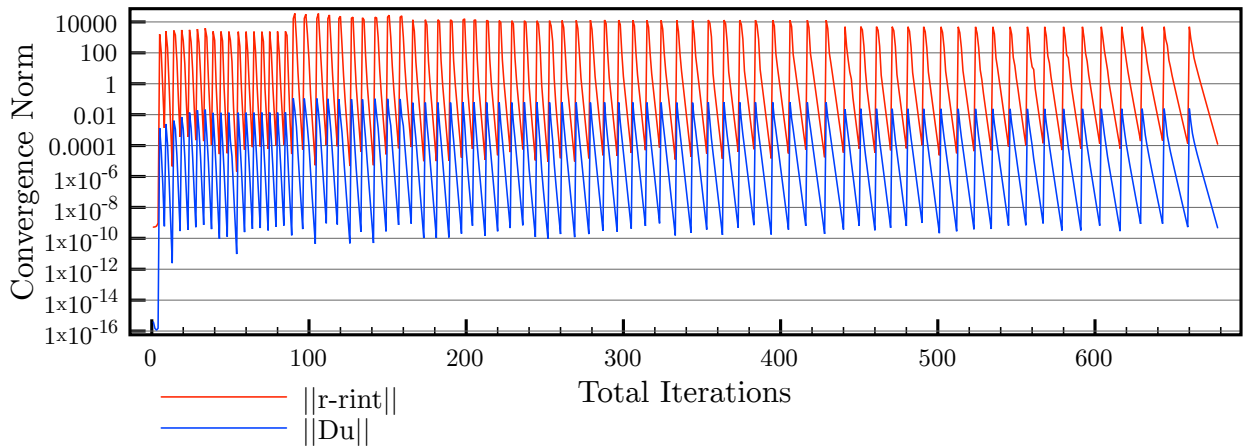

To further study convergence, the same geometry as the notched specimen was modeled without a notch. This cylindrical geometry was composed of an entirely regular node pattern, containing a single out-of-place node. This out-of-place node serves as a location for damage nucleation, in order to avoid uniformly distributed damage. It was observed that, for this geometry, convergence of the kinematic hardening model is excellent; global convergence was regularly achieved within three iterations. Unfortunately, sufficiently high levels of damage ($D = 0.4$) could only be obtained for displacements of up to $u = 9\,\text{mm}$, well out of the range of infinitesimal strain theory.

The author therefore cannot completely verify the correct implementation of the kinematic hardening model. Further investigation must be carried out to insure that the peculiarities observed are fully understood. An attempt was made to point out the difficulties arising from the use of VOIGT notation in section 4.3. It is possible that errors can be found to be related to such difficulties.

Figure 4.4: Global equilibrium iterations vs. vertical displacement of upper edge. Calculation of model with 121 nodes, 32 elements. Both models were subject to the same load increments for upper edge displacement below 0.57 mm.

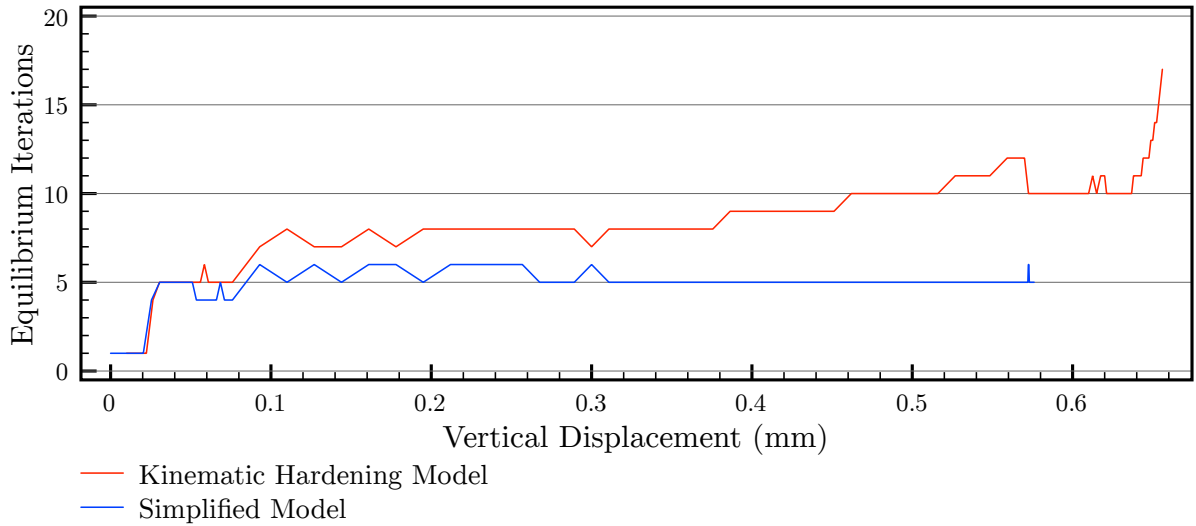

# 5   Conclusions

An attempt was made to give a complete account of the relations constituting LEMAITRE's ductile damage model by presenting relevant parts of the basic theory of plasticity and continuum damage mechanics. The underlying physical causes for plasticity and damage from a standpoint of materials science were reviewed. Furthermore, the steps needed to implement the model within standard nonlinear finite element code were outlined. In particular, the derivatives needed for the Jacobian matrix were given, and an attempt was made to provide a simple step-by-step derivation of the consistent elastoplastic tangent modulus. Such an implementation written in MATLAB was developed. Simplifications arising from the absence of kinematic hardening were presented and also implemented in computer code.

Within the realm of the model's limitations, both implementations were found to exhibit the physical behavior expected from theory. The simplified model was verified by comparison with results from literature and showed quadratic levels of convergence, indicating correct implementation. Convergence of the model including kinematic hardening was found to deteriorate with damage, as is described in literature (cf. [11]). Although it is possible that the levels of convergence deterioration observed arise from particularities of the geometry studied, it was concluded that further investigation is needed to sufficiently verify the model. Geometries found in literature (cf. [3]) should therefore be studied in order to ensure that convergence is within acceptable limits.

# A  Figures Relating to Computations

Figure A.1: Cylindrical notched specimen: geometry and meshes (left: 32 elements, 121 nodes; right: 512 elements, 1633 nodes)

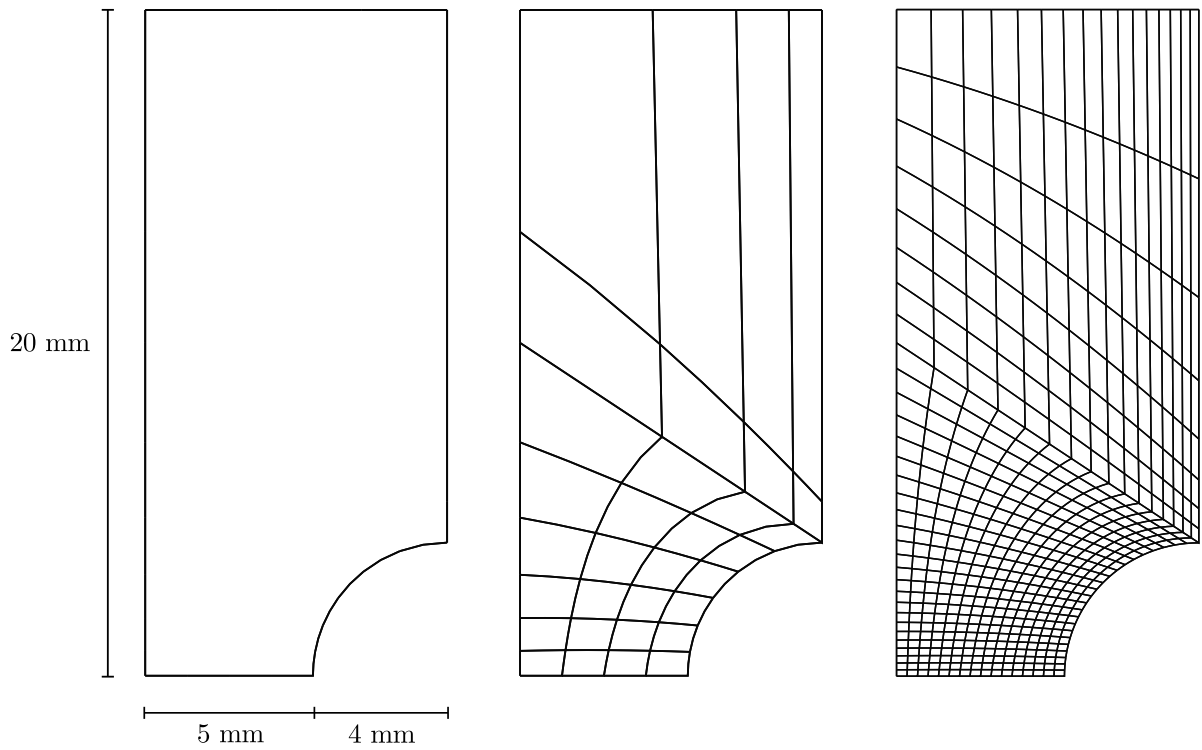

Figure A.2: Damage at maximum damaged Gauss point (element 9, Gauss point 1) vs. vertical displacement of upper edge. Calculation of model with 121 nodes, 32 elements, results may be inaccurate.

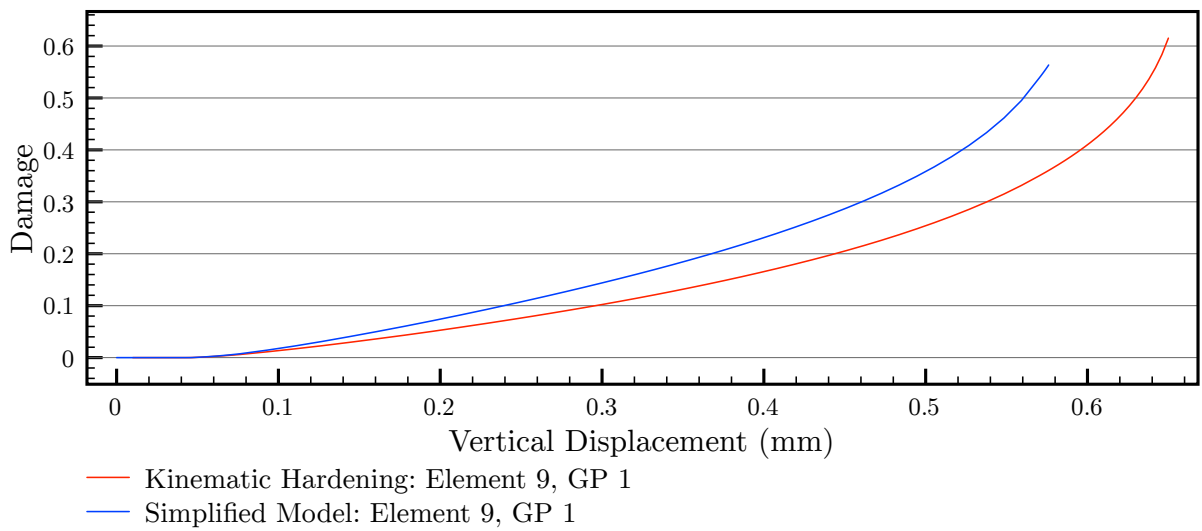

## A.1 Lemaitre's Simplified Model

Figure A.3: Comparison of results with the reference work [11]. Damage at maximum damaged GAUSS point (element 9, GAUSS point 1) vs. vertical displacement of upper edge. Calculation of model with 121 nodes, 32 elements, results may be inaccurate.

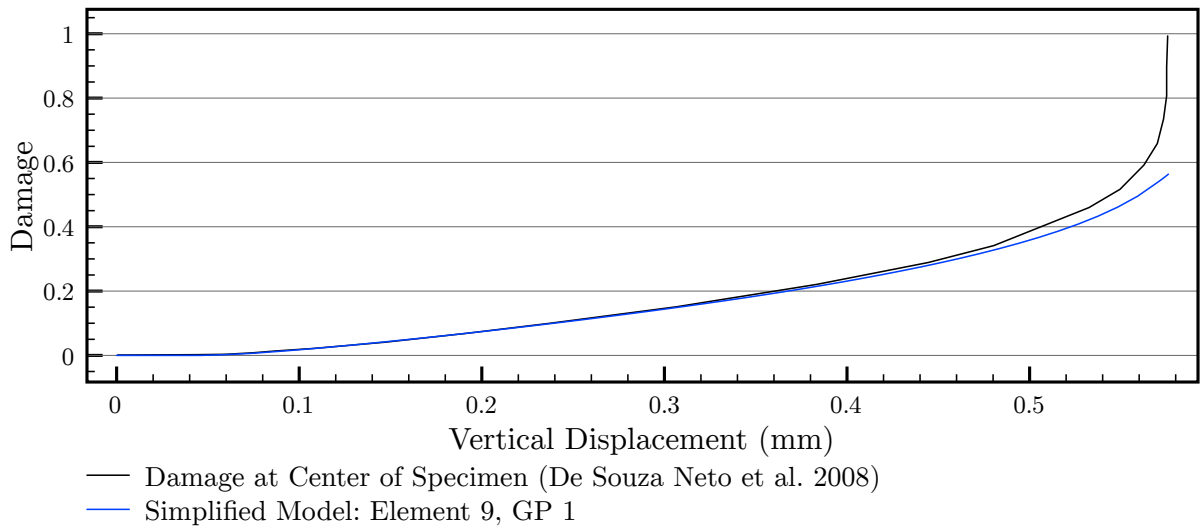

Figure A.4: Observed true stress $q(\boldsymbol{\sigma}_d)$ for Lemaitre's simplified model and true stresses $(1 - D)\kappa$ and $(1 - D)\sigma_{y0}$ at maximum damaged Gauss point (element 9, Gauss point 1) vs. vertical displacement of upper edge. Calculation of model with 121 nodes, 32 elements, results may be inaccurate.

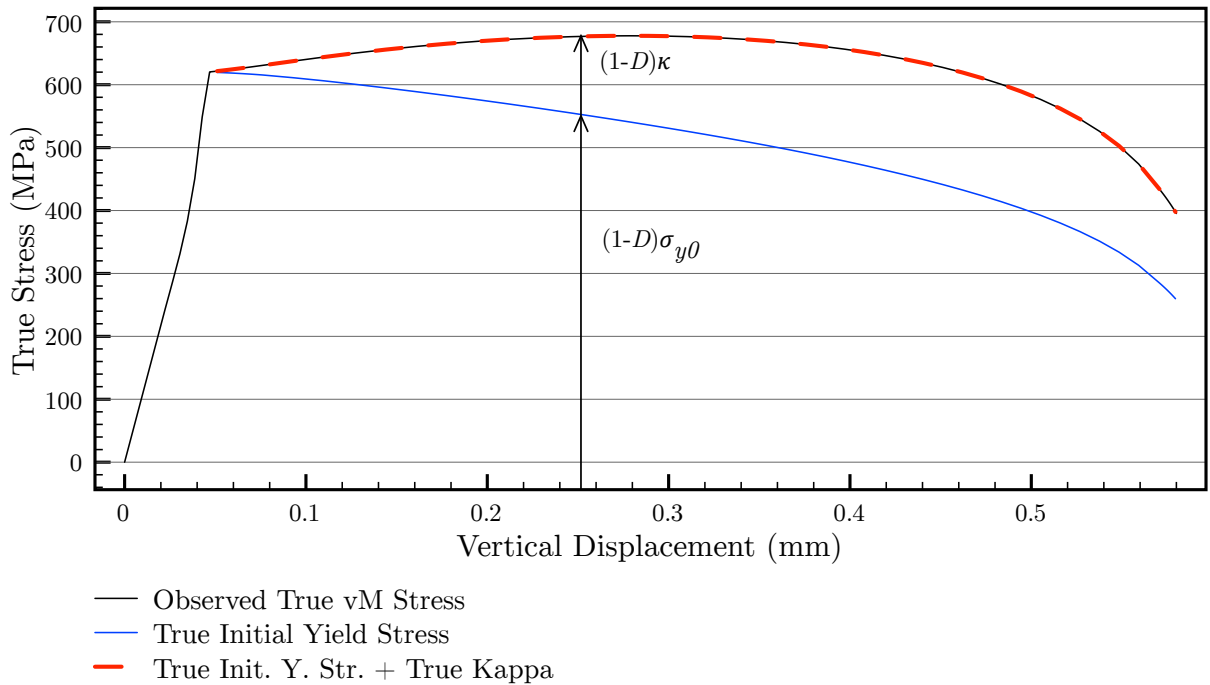

Figure A.5: Lemaitre's simplified model: nodal damage (512 elements, 1633 nodes)

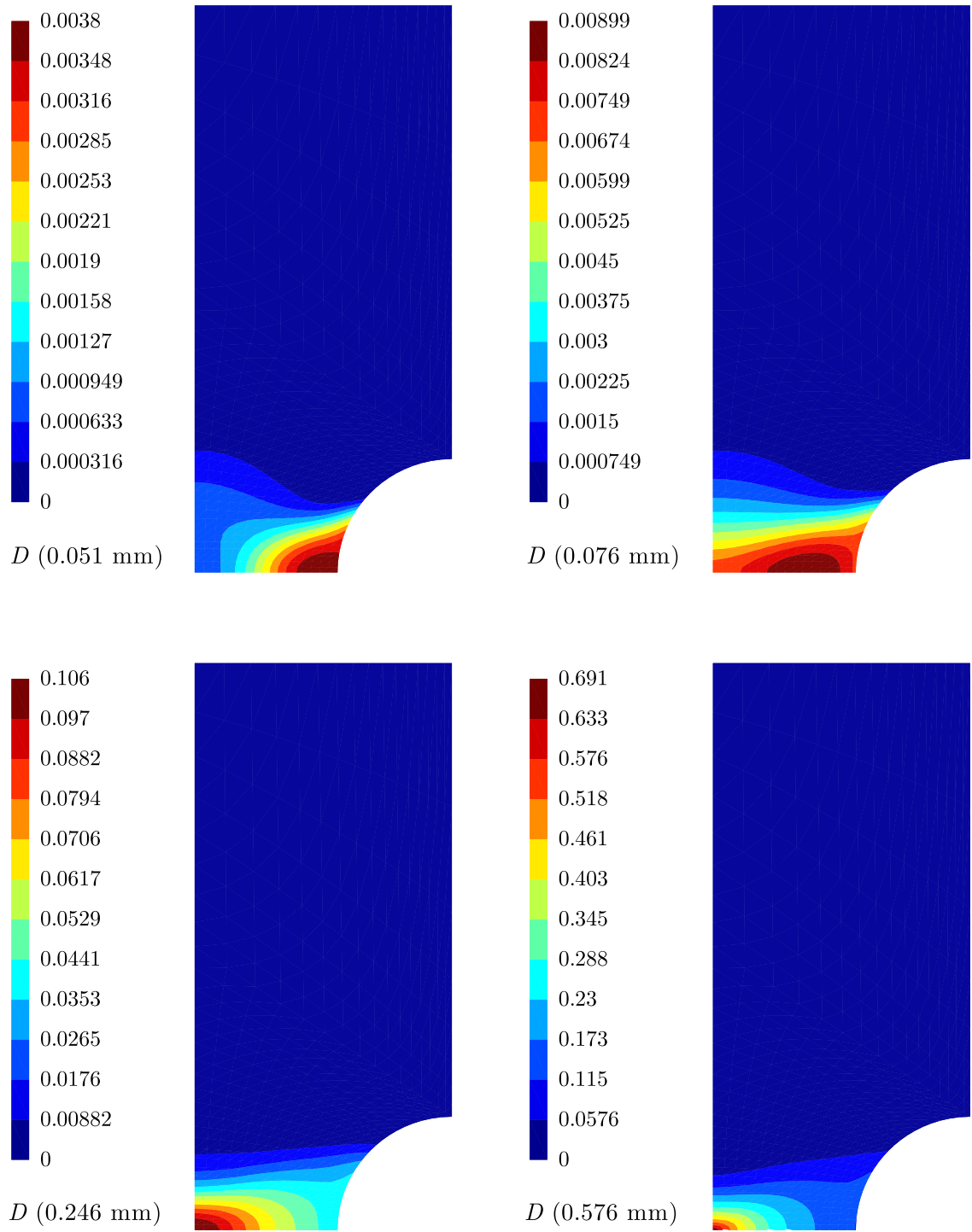

Figure A.6: Lemaitre's simplified model: damage plots from de Souza Neto et al. (2008) [11].

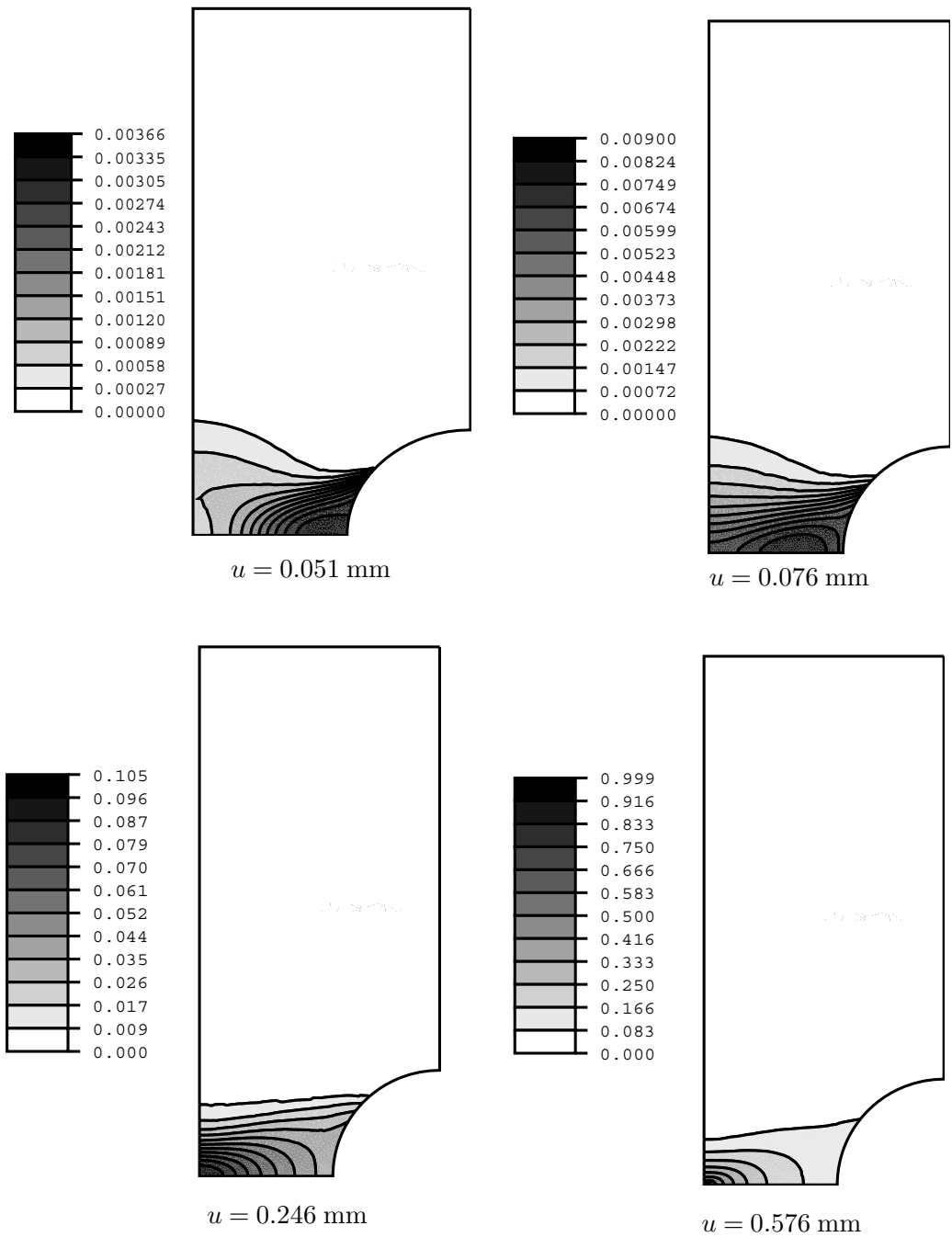

Figure A.7: Lemaitre's simplified model: stress triaxiality (512 elements, 1633 nodes)

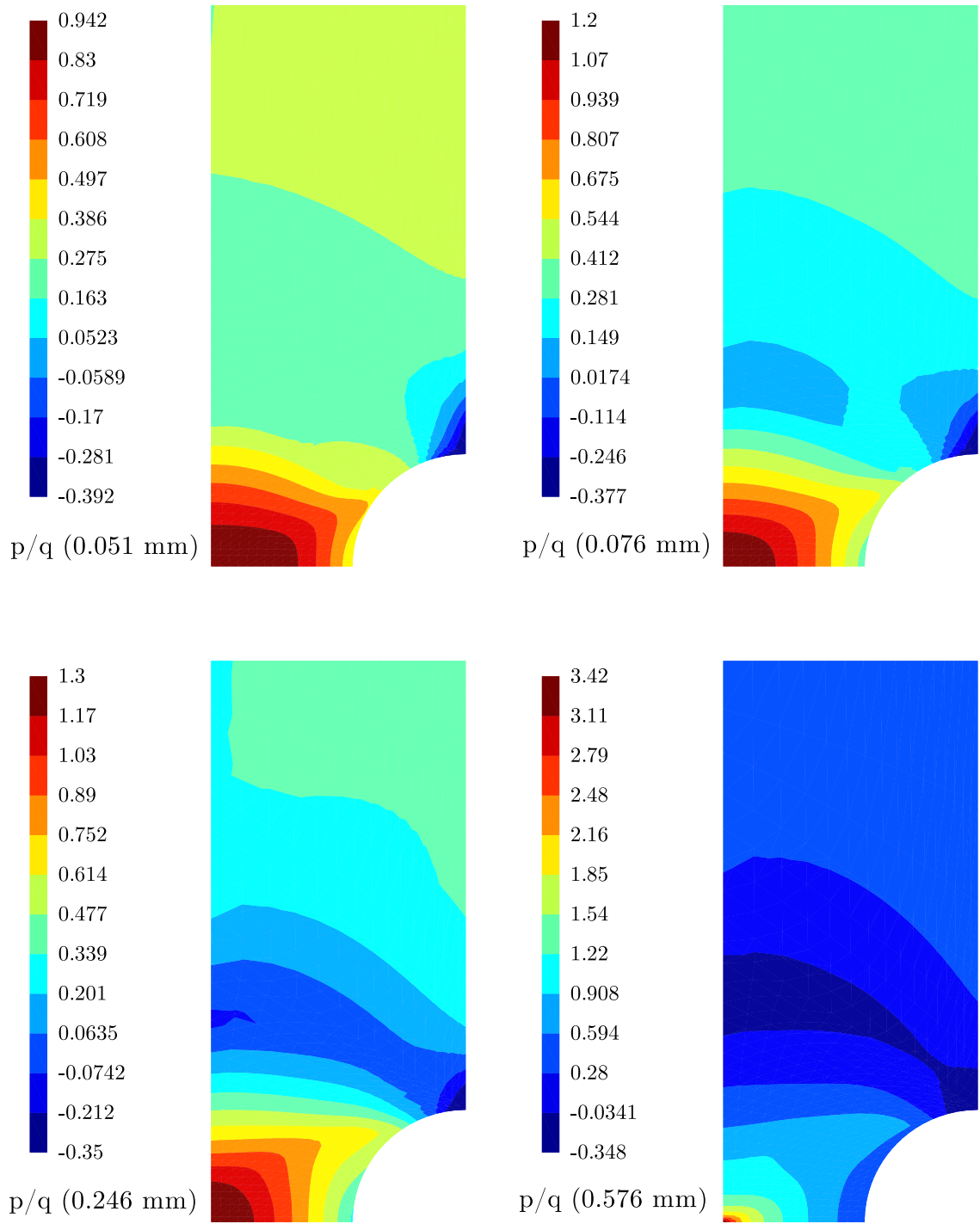

## A.2   Lemaitre's Model Including Kinematic Hardening

Figure A.8: Observed true stress $q(\boldsymbol{\sigma}_d)$ for LEMAITRE's model including kinematic hardening and true stresses $(1-D)\kappa$, $(1-D)\sigma_{y0}$, and $q(\boldsymbol{\beta})$ at maximum damaged GAUSS point (element 9, GAUSS point 1) vs. vertical displacement of upper edge. Calculation of model with 121 nodes, 32 elements, results may be inaccurate.

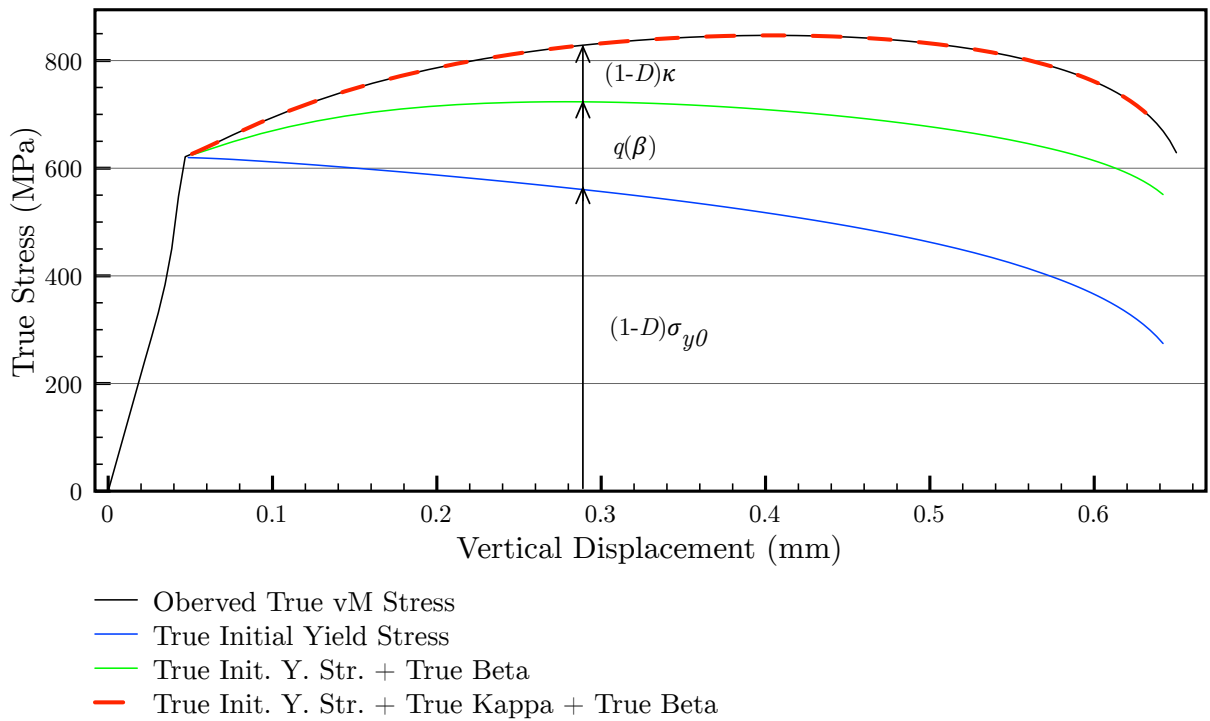

Oberved True vM Stress
True Initial Yield Stress
True Init. Y. Str. + True Beta
True Init. Y. Str. + True Kappa + True Beta

Figure A.9: Observed true VON MISES stress $q(\boldsymbol{\sigma}_d)$ at maximum damaged GAUSS point (element 9, GAUSS point 1) vs. vertical displacement of upper edge. Cyclic loading, $R_u = -1$, $u = 0.076$ mm. Calculation of model with 121 nodes, 32 elements, results may be inaccurate.

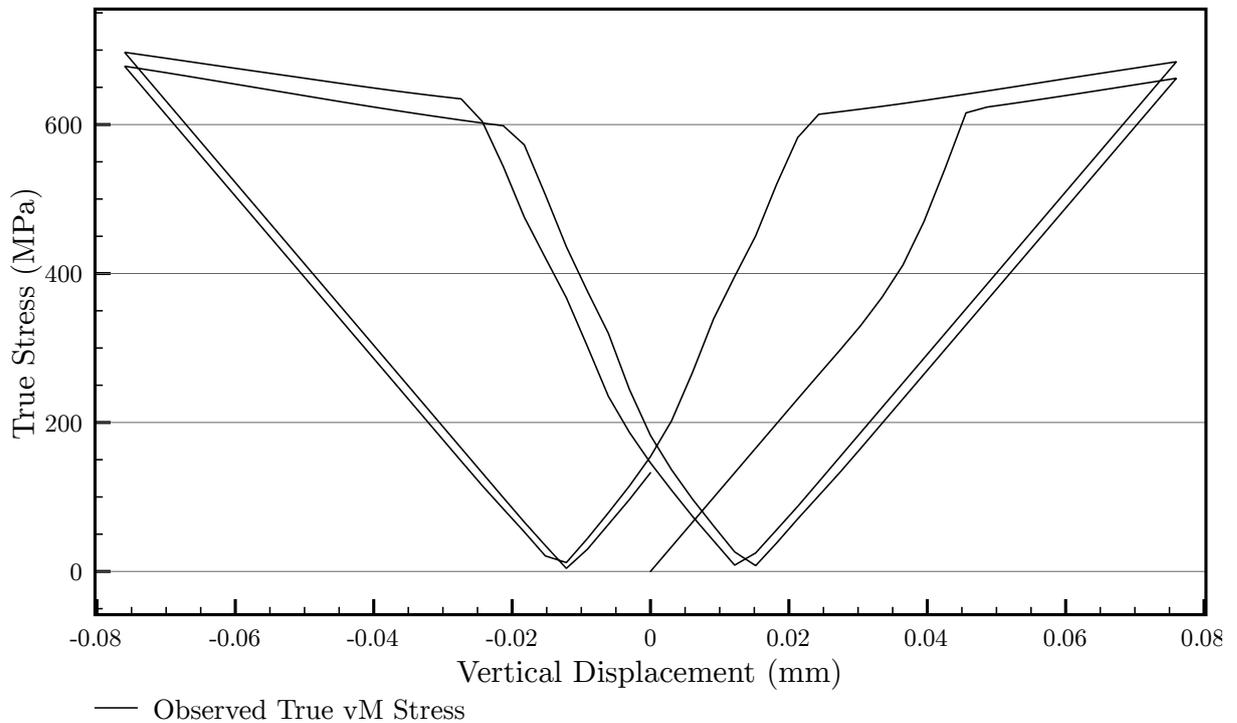

———  Observed True vM Stress

Figure A.10: LEMAITRE's model including kinematic hardening: nodal damage (512 elements, 1633 nodes)

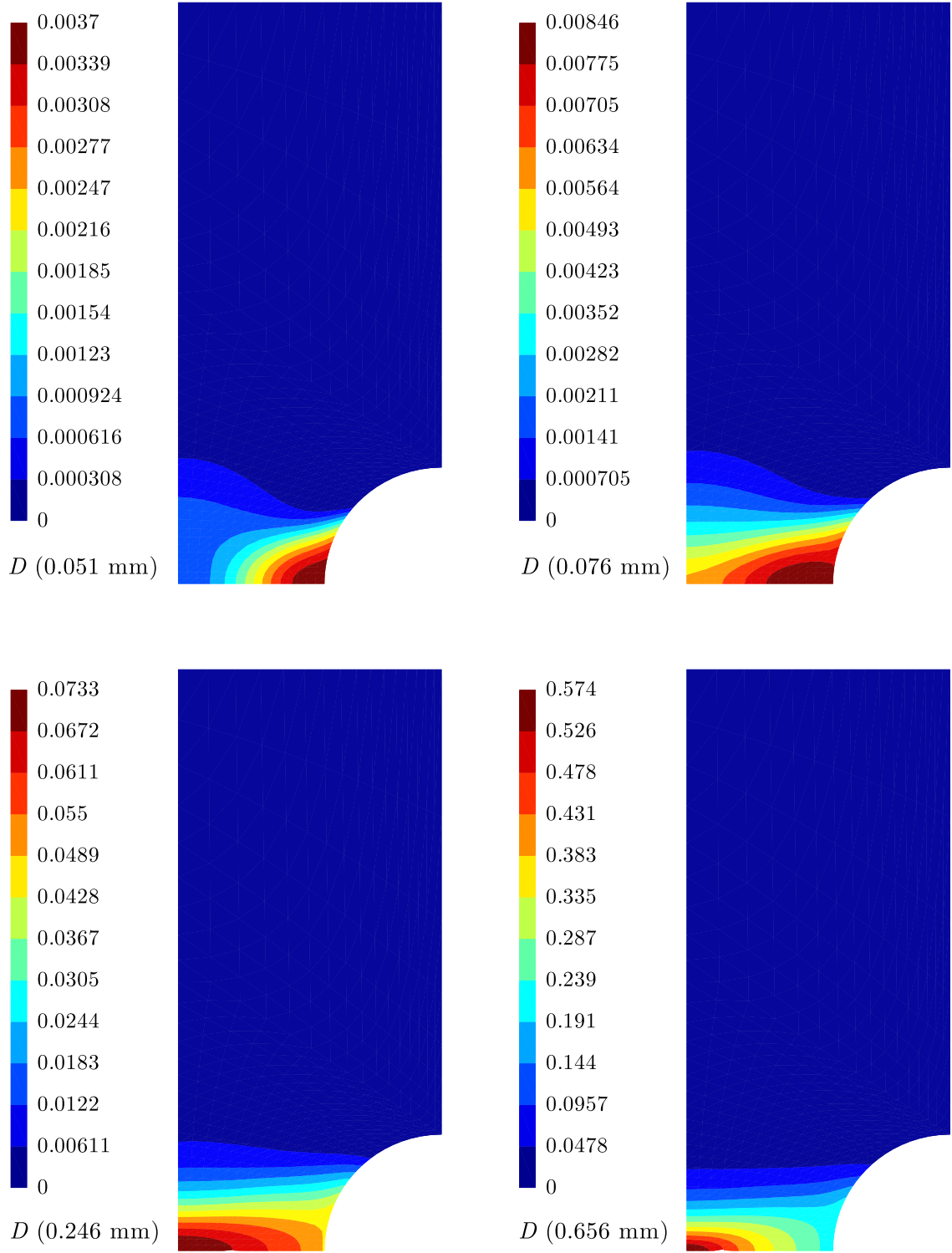

Figure A.11: LEMAITRE's model including kinematic hardening: stress triaxiality (512 elements, 1633 nodes)

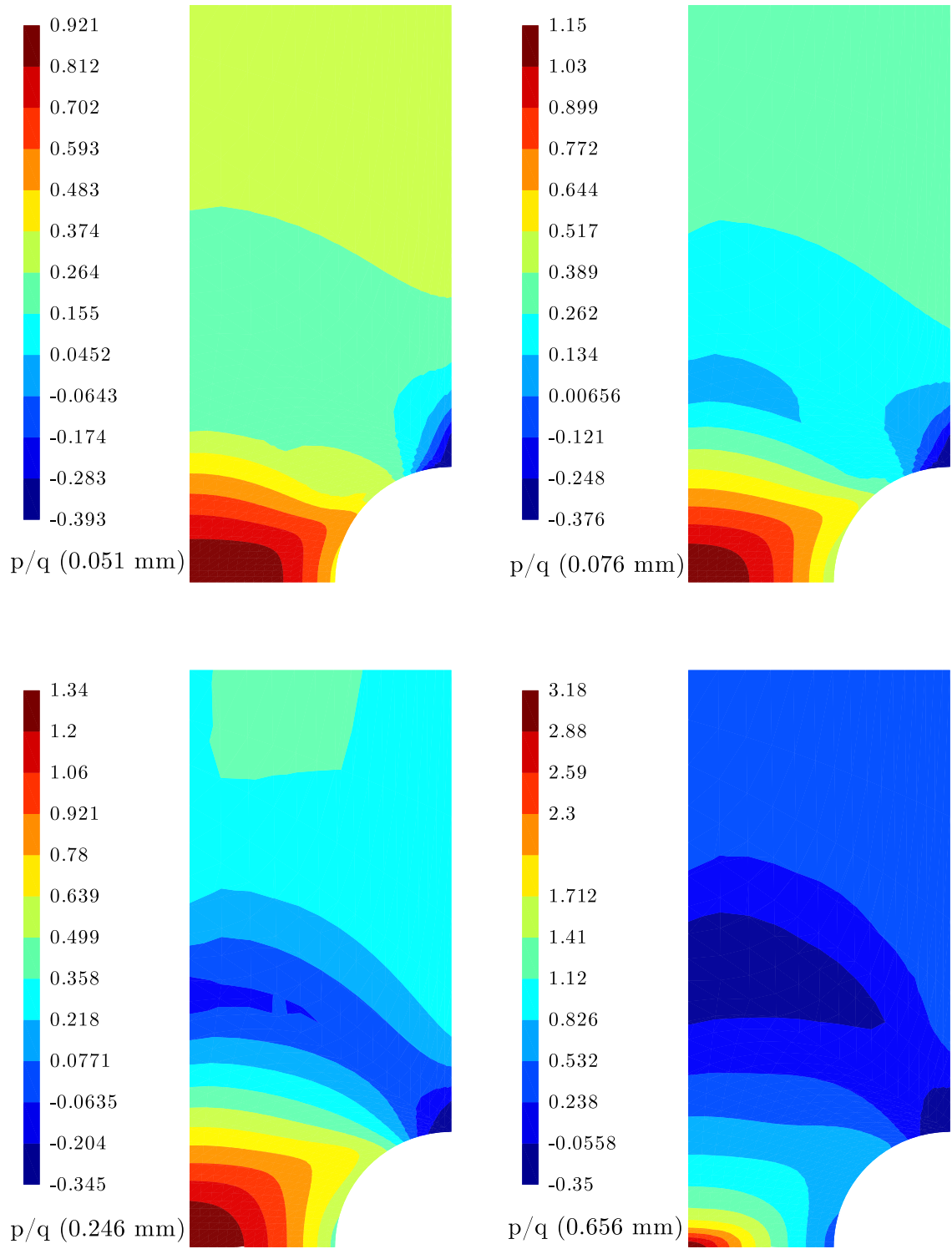

Figure A.12: LEMAITRE's model including kinematic hardening: $\beta_{11}$ (MPa) (512 elements, 1633 nodes)

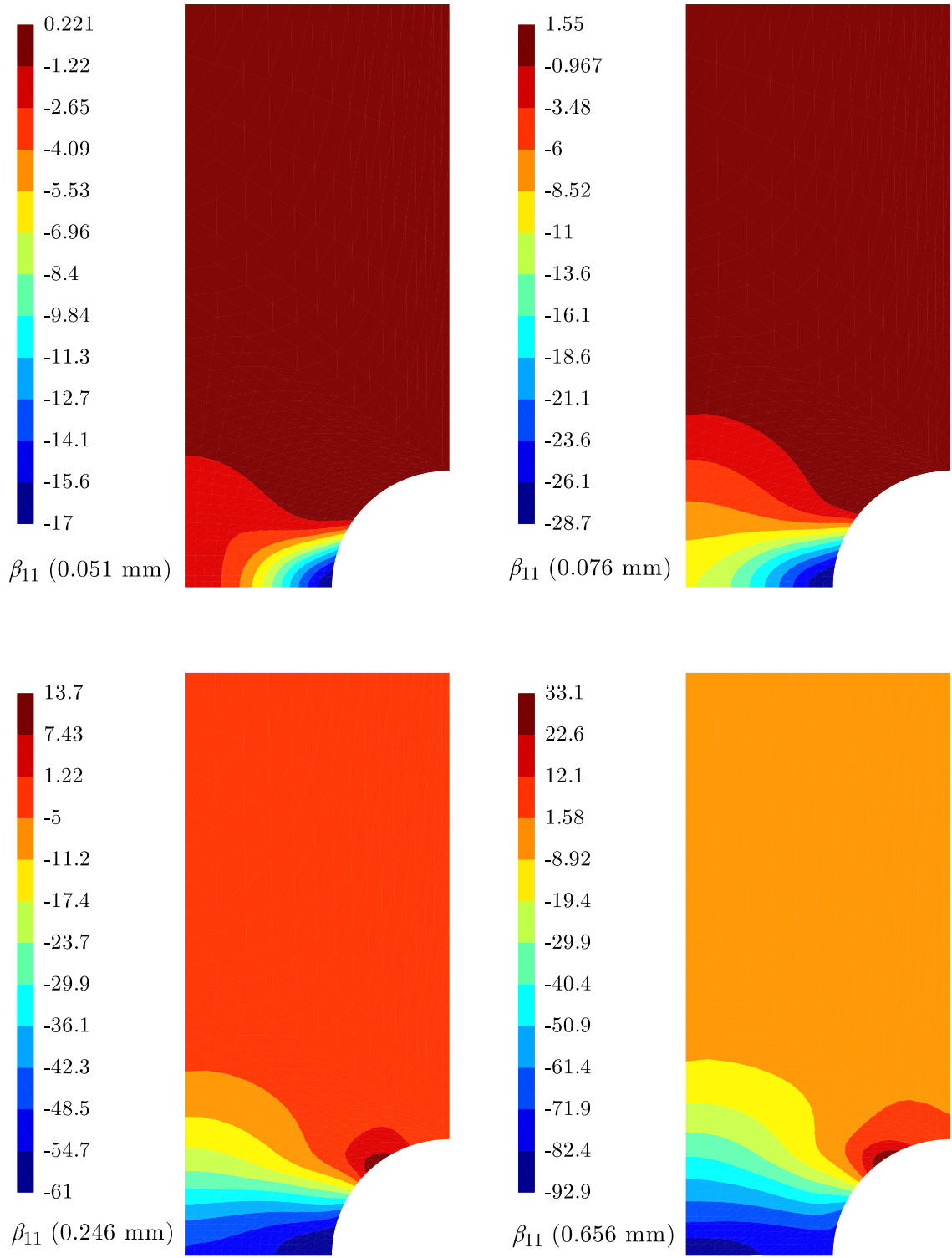

Figure A.13: LEMAITRE's model including kinematic hardening: $\beta_{22}$ (MPa) (512 elements, 1633 nodes)

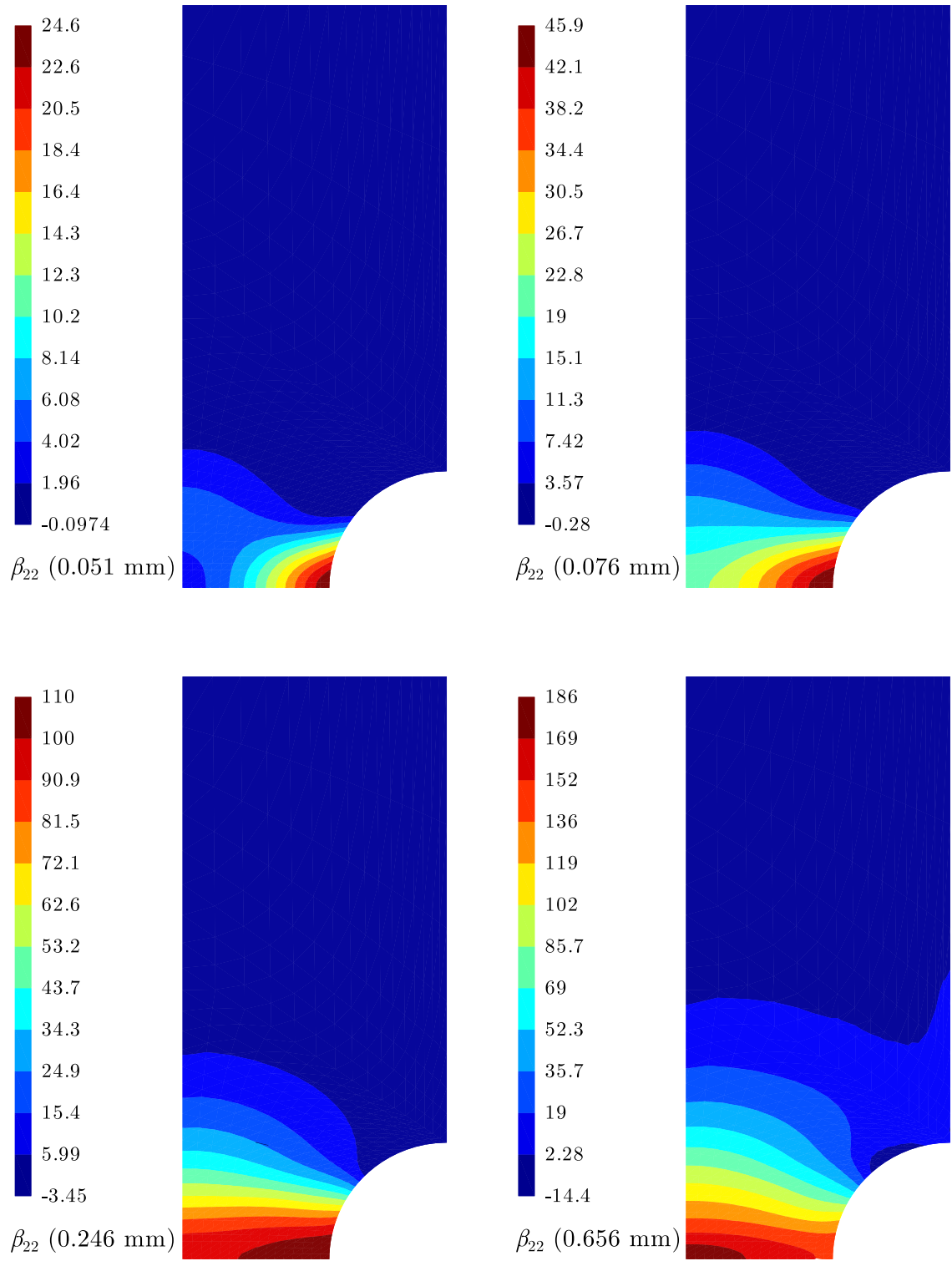

Figure A.14: LEMAITRE's model including kinematic hardening: $\beta_{33}$ (MPa) (512 elements, 1633 nodes)

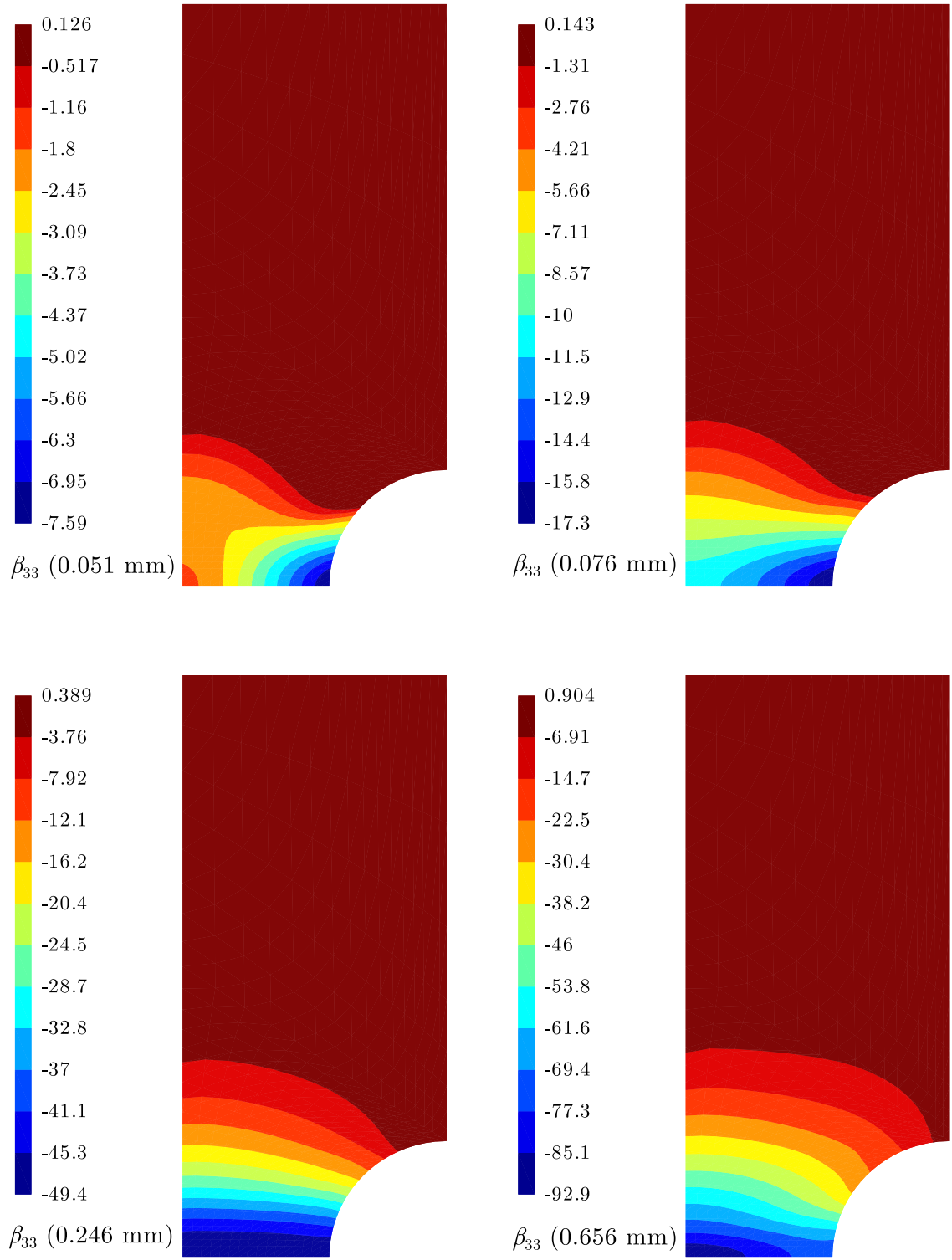

Figure A.15: Lemaitre's model including kinematic hardening: $\beta_{12}$ (MPa) (512 elements, 1633 nodes)

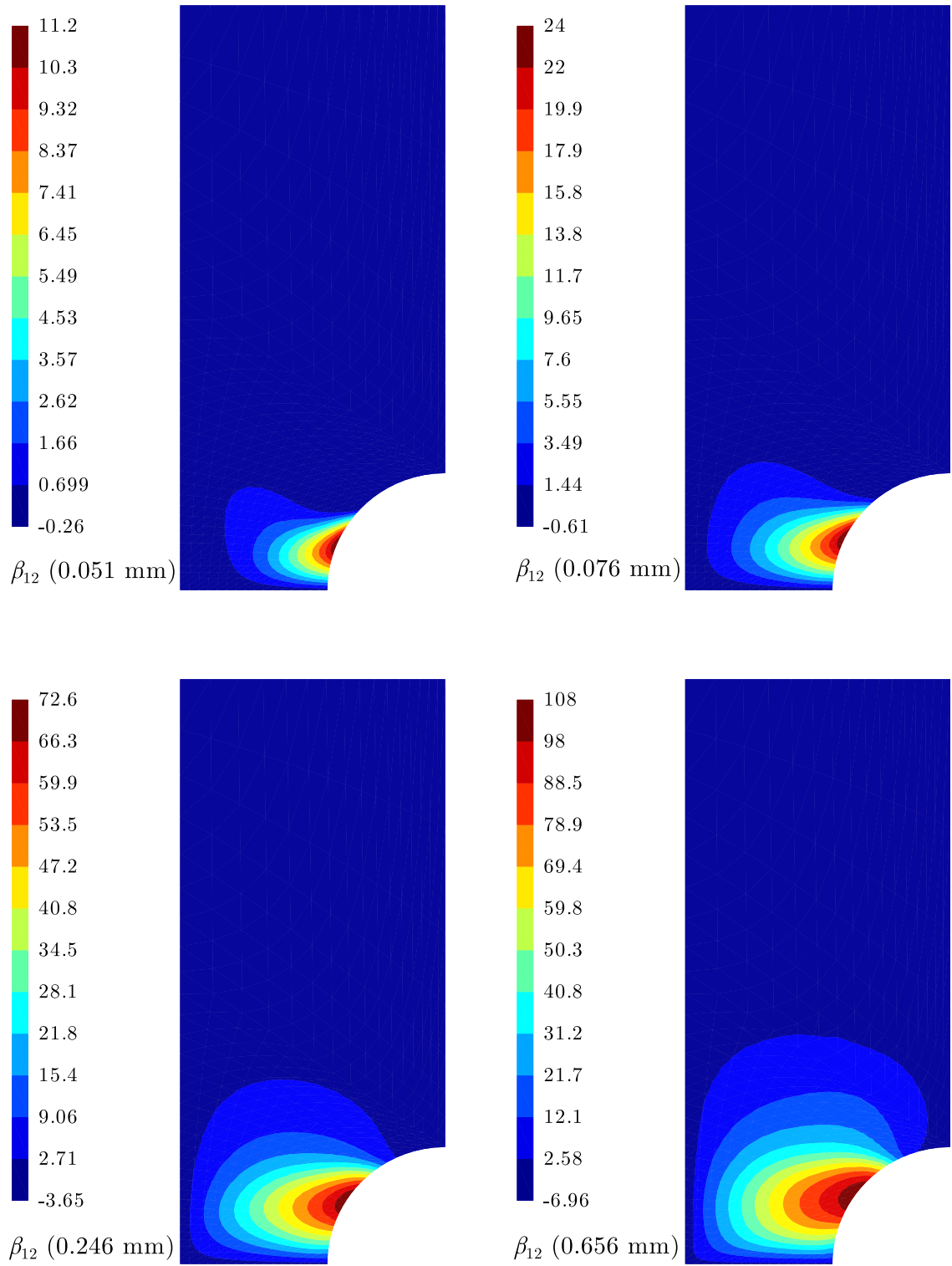

# B  Source Code

## B.1  Lemaitre's Simplified Damage Model

```matlab
1  % Implementation of Lemaitre's Simplified Damage Model
2  % For 3D, 2D and axisymmetric problems
3  % Based on the algorithm implemented in the FORTRAN program
4  % HYPLAS developed by De Souza Neto et al. (2008)
5
6  % Bachelor Thesis:
7  % "A Finite Element Implementation of a Ductile Damage Model for Small Strains"
8  % Author: Robert Lee Gates
9
10 % Institute of Mechanics and Computational Mechanics
11 % University of Hannover, Germany
12
13 % Email: robert.gates@gmail.com
14
15
16 % MATLAB function
17 % Returns material tangent for Lemaitre's simplified model
18 function [ D ] = plast_lemaitre_simple(i, ig)
19
20 %% Parameters
21 global mat;
22 global eldat;
23
24 tol = 10^(-7);
25 dim = eldat.compute_dim;
26 if dim == 3
27    % array for hydraulic component retrieval
28    Ih = [1;1;1;0;0;0];
29    Ih2 = [0;0;0;1;1;1];
30    % 2nd order identity tensor in voigt notation
31    I = [1;1;1;0;0;0];
32    % conversion arrays
33    enToPhys = [1;1;1;.5;.5;.5];
34    physToEn = [1;1;1;2;2;2];
35    % number of unique stresses
36    ncomp = 6;
37    % linear elastic material tangent
```

```
38   Ce = lin_material(i,0);
39
40   % 2nd order identity tensor in Voigt notation
41   I = [1; 1; 1; 0; 0; 0];
42 elseif dim == 2
43   % array for hydraulic component retrieval
44   Ih = [1;1;0;1];
45   Ih2 = [0;0;1;0];
46   % conversion arrays
47   enToPhys = [1;1;.5;1];
48   physToEn = [1;1;2;1];
49   % number of unique stresses
50   ncomp = 4;
51   % linear elastic material tangent
52   Ce = lin_material(i,1);
53
54   % 2nd order identity tensor in Voigt notation
55   I = [1; 1; 0; 1];
56 end
57
58 % tensor product of two 2nd order identity tensors in Voigt notation
59 IxI = I*I';
60
61 % 4th order symmetric projection tensor in Voigt notation
62 Is = 0.5.*(diag(I) + eye( ncomp));
63
64 % deviatoric projection tensor in Voigt notation
65 Id = Is - (1./3) .* IxI;
66
67 % material parameters (see input_lemaitre_lmat.m)
68 matnr = eldat.mat(i);
69 sig_y0 = mat.sig0( matnr );          % initial yield stress
70 G = mat.G( matnr );                  % shear modulus
71 K = mat.K( matnr );                  % bulk modulus
72 r = mat.r( matnr );                  % isotropic hardening
73 s = mat.s( matnr );                  % isotropic hardening
74 Rinf = mat.Rinf( matnr);             % isotropic hardening
75 gamma = mat.gamma(matnr);            % isotropic hardening
76
77 % some factors
78 K2 = 2.*K;
79 G2 = 2.*G;
80 G3 = 3.*G;
81 G6 = 6.*G;
82
83 %% Get Strain, Damage, Hardening
84 eps = eldat.epsilon(i, :, ig)';      % total strain
85 eps_pl = eldat.eps_pl(i, :, ig)';    % plastic strain
86 Dam0 = eldat.damage(i, ig);          % initial damage
87 Int0 = 1 - Dam0;                     % initial integrity
88 R0 = eldat.R(i, ig);                 % initial hardening internal variable
89
90 %% Trial State
91 eps_e_tr = eps - eps_pl;             % elastic trial strain
92 eps_e_hyd_tr = sum(eps_e_tr.*Ih);    % hydrostatic strain
```

```matlab
93  sig_hyd_tr = K.*eps_e_hyd_tr;                          % hydrostatic effective stress
94  eps_e_dev_tr = eps_e_tr - (1./3).*eps_e_hyd_tr.* Ih;   % deviatoric strain
95  % convert engineering shear strain to physical shear strain
96  eps_e_dev_tr = eps_e_dev_tr .* enToPhys;
97
98
99  % compute effective trial von Mises stress
100 temp = eps_e_dev_tr.^2;
101 J_2 = G2.^2 .* ( .5*sum(temp.*Ih) + sum(temp.*Ih2) );
102 q_tr = sqrt(3*J_2);
103
104 % compute yield stress
105 fsig0 = sig_y0 + Rinf.*(1-exp(-gamma.*R0));
106
107
108 Phi = q_tr - fsig0;
109
110 %% Check if Yield Criterion is met
111 if Phi >= 0
112
113   % return mapping
114   % initial guess for the plastic multiplier
115   plasticMult = Int0.*Phi./(3.*G);
116
117   % inital guess for the hardening variable
118   R = R0 + plasticMult;
119
120   % initial values
121   norm_F = 1;
122   sig_hyd_tr2 = sig_hyd_tr.^2;
123
124   % Newton Raphson iteration for finding the true plastic multiplier
125   iter = 0;
126   maxiter = 100;
127   while norm_F >= tol && iter <= maxiter
128
129     if iter == maxiter
130       disp('Fatal: return mapping reached maximum iterations!');
131       break;
132     end
133
134     % current yield stress
135     fsig = sig_y0 + 3300.*(1-exp(-.4.*R));
136
137     % integrity & strain energy release rate function
138     f1 = (G3 ./ (q_tr - fsig));
139     Int = f1 .* plasticMult;
140     Y = -(fsig.^2)./G6 - (sig_hyd_tr2)./K2;
141     f2 = -Y./r;
142
143     % compute residual
144     F = Int - Int0 + (-Y./r).^s ./f1;
145     norm_F = abs(F);
146
147     % derivatives
```

```
148      dfsig = 1320 .* exp(-.4.*(R));
149      dY = -(fsig.*dfsig) ./ G3;
150
151      % residual derivative
152
153      f = f1 + f1.*plasticMult.*dfsig./(q_tr - fsig) - ...
154        (dfsig./G3).*f2.^s - (s.*dY./(f1.*r)).*f2.^(s-1);
155
156      % get next plastic multiplier
157      plasticMult = plasticMult - F./f;
158
159      % update hardening variable
160      R = R0 + plasticMult;
161
162      iter = iter + 1;
163
164  end
165
166  % having now received the true plastic multiplier, update
167  % effective yield stress
168  fsig = sig_y0 + Rinf.*(1-exp(-gamma.*R));
169  % hardening slope
170  dfsig = gamma .* Rinf .* exp(-gamma.*R);
171  % integrity
172  f1 = (G3 ./ (q_tr - fsig));
173  Int = f1 .* plasticMult;
174  dInt = (G3+Int.*dfsig)./(q_tr-fsig);
175  % strain energy release rate
176  Y = -(fsig.^2)./G6 - (sig_hyd_tr2)./K2;
177  dY = -(fsig.*dfsig) ./ G3;
178  f2 = -Y./r;
179  % residual derivative
180  f = f1 + f1.*plasticMult.*dfsig./(q_tr - fsig) - ...
181      (dfsig./G3).*f2.^s - (s.*dY./(f1.*r)).*f2.^(s-1);
182
183  % check if NR yielded an acceptable damage variable
184  if(Int < 10^-20)
185    disp('GP integrity too small!');
186  end
187
188  % update damage
189  Dam = 1-Int;
190
191  % update true von Mises stress
192  q = Int .* fsig;
193
194  % update true stresses, strains
195  sig_hyd = Int .* sig_hyd_tr;
196  sig_dev = G2.*(q./q_tr).*eps_e_dev_tr;
197  sig = sig_dev + sig_hyd .* Ih;
198
199  % plastic corrector
200  plCor = G3.*plasticMult./(Int .* q_tr);
201  % restore engineering strain and update total elastic strain
202  eps_e = (1-plCor) .* eps_e_dev_tr .* physToEn + (1./3).*eps_e_hyd_tr .* Ih;
```

```
203    % increase in plastic strain
204    deps_pl = eps_e_tr - eps_e;
205
206    % update model
207    eldat.sigma(i, :,ig) = sig;
208    eldat.eps_pl_n(i,:, ig) = eps_pl + deps_pl;
209    eldat.damage_n(i, ig) = Dam;
210    eldat.R_n(i, ig) = R;
211    eldat.q_n(i, ig) = q;
212    eldat.triax_n(i, ig) = sig_hyd./q;
213    eldat.sigy_n(i, ig) = fsig;
214
215    % norm of deviatoric stress
216    snorm = sqrt(sum(physToEn.*(sig_dev.^2)));
217
218    % compute coefficients for elastoplastic tangent
219    f3 = q_tr - fsig;
220    a1 = (1./f).*( Int./f3 - (1./G3).*f2.^s );
221    a2 = -s.*sig_hyd_tr.*f3./( G3.*r.*K.*f ).*f2.^(s-1);
222    a3 = a2.*dInt;
223    a4 = a1.*dInt - Int./f3;
224    a = G2.*Int*fsig./q_tr;
225    b = G2.*( a1.*dfsig.*Int + a4.*fsig - Int.*fsig./q_tr );
226    b = b./(snorm.^2);
227    c = K.*sqrt(2./3).*( a2.*dfsig.*Int + a3.*fsig );
228    c = c./snorm;
229    d = G2.*sqrt(3./2).*sig_hyd_tr.*a4;
230    d = d./snorm;
231    e = K.*(Int + a3.*sig_hyd_tr);
232
233    % Consistent Elastoplastic Tangent Operator
234    % (minor symmetric, not major symmetric)
235    D = a.*Id + b.*sig_dev*sig_dev' + c.*sig_dev*I' + d.*I*sig_dev' + e.*IxI;
236
237
238 % If yield criterion is not met, material is behaving elastically
239 else
240    D = Int0.*Ce;
241
242    % true stress components
243    sig_hyd = Int0 .* sig_hyd_tr;
244    sig_dev = G2 .* Int0 .* eps_e_dev_tr;
245    % true stress
246    sig = sig_dev + sig_hyd .* Ih;
247    % update model
248    eldat.sigma(i, :,ig) = sig;
249    % plastic strain
250    eldat.eps_pl_n(i,:, ig) = eps_pl;
251    % initial damage
252    eldat.damage_n(i, ig) = Dam0;
253    % isotropic hardening internal variable
254    eldat.R_n(i, ig) = R0;
255    % true vM stress
256    eldat.q_n(i, ig) = Int0.*q_tr;
257    % stress triaxiality
```

```
258   eldat.triax_n(i, ig) = sig_hyd./(Int0.*q_tr);
259   % effective yield stress
260   eldat.sigy_n(i, ig) = fsig0;
261 end
```

## B.2   Lemaitre's Damage Model Including Kinematic Hardening

```
1  % Implementation of Lemaitre's Damage Model Including Kinematic Hardening
2  % For 3D, 2D and axisymmetric problems
3
4  % Bachelor Thesis:
5  % "A Finite Element Implementation of a Ductile Damage Model for Small Strains"
6  % Author: Robert Lee Gates
7
8  % Institute of Mechanics and Computational Mechanics
9  % University of Hannover, Germany
10
11 % Email: robert.gates@gmail.com
12
13
14 % MATLAB function
15 % Returns material tangent for Lemaitre's model
16 function [ Dp ] = plast_lemaitre_kinematic(i, ig)
17
18 %% Parameters
19 global mat;
20 global eldat;
21 global inr;
22
23 tol = 10^(-7);
24 dim = eldat.compute_dim;
25 if dim == 3
26   % array for hydraulic component retrieval
27   Ih = [1;1;1;0;0;0];
28   % conversion arrays
29   physToEn = [1;1;1;2;2;2];
30   % number of unique stresses
31   ncomp = 6;
32   % linear elastic material tangent
33   Ce = lin_material(i,0);
34
35   % 2nd order identity tensor in Voigt notation
36   I = [1 1 1 0 0 0]';
37 elseif dim == 2
38   % array for hydraulic component retrieval
39   Ih = [1;1;0;1];
40   % conversion arrays
41   physToEn = [1;1;2;1];
42   % number of unique stresses
43   ncomp = 4;
```

```matlab
44   % linear elastic material tangent
45   Ce = lin_material(i,1);
46
47   % 2nd order identity tensor in Voigt notation
48   I = [1 1 0 1]';
49 else
50   error('Wrong dimensions!');
51 end
52
53 % tensor product of two 2nd order identity tensors in Voigt notation
54 IxI = I*I';
55
56 % 4th order symmetric projection tensor resulting from derivative
57 % in Voigt notation
58 Is1 = 0.5.*(diag(I) + eye(ncomp));
59
60 % 4th order deviatoric projection tensor resulting from derivative
61 % in Voigt notation
62 Id1 = Is1 - (1./3) .* IxI;
63
64 if dim == 2
65   Is1(3,3) = 1;
66   Id1(3,3) = 1;
67 elseif dim == 3;
68   Is1(4,4) = 1;
69   Is1(5,5) = 1;
70   Is1(6,6) = 1;
71   Id1(4,4) = 1;
72   Id1(5,5) = 1;
73   Id1(6,6) = 1;
74 end
75
76 % material parameters (see input_lemaitre_lmat.m)
77 matnr = eldat.mat(i);
78 sig_y0 = mat.sig0( matnr );        % initial yield stress
79 G = mat.G(matnr);                  % shear modulus
80 K = mat.K(matnr);                  % bulk modulus
81 r = mat.r(matnr);                  % isotropic hardening
82 s = mat.s(matnr);                  % isotropic hardening
83 a = mat.a(matnr);                  % kinematic hardening
84 b = mat.b(matnr);                  % kinematic hardening
85 Rinf = mat.Rinf( matnr);           % isotropic hardening
86 gamma = mat.gamma(matnr);          % isotropic hardening
87
88 G2 = 2.*G;
89
90 %% Get Strain, Damage, Hardening, Back Stress
91 eps = eldat.epsilon(i, :, ig)';    % total strain
92 eps_pl = eldat.eps_pl(i, :, ig)';  % plastic trial strain
93 D0 = eldat.damage(i, ig);          % initial damage
94 Int0 = 1 - D0;                     % initial integrity
95 R0 = eldat.R(i, ig);               % initial hardening internal variable
96 beta0 = eldat.beta(i, :, ig)';     % initial TRUE back stress
97
98 %% Trial State
```

```
99  eps_e_tr = eps - eps_pl;                     % elastic trial strain
100 sig_tr = Int0.*Ce*eps_e_tr;                   % elastic trial TRUE stress
101
102 % compute TRUE hydrostatic and deviatoric trial stress
103 % eps_hyd = trace(eps)
104 % sig = K*eps_hyd + 2G*eps_dev
105 % distortional or deviatoric stress
106 sig_hyd_tr = sum(sig_tr.*Ih)./3;
107 % hydrostatic stress
108 sig_dev_tr = sig_tr - sig_hyd_tr.*I;
109
110 % compute relative trial stress
111 rel_tr = sig_dev_tr - beta0;
112
113 % compute second invariant of relative stress
114 J_2 = sum((rel_tr.^2).*physToEn)./2;
115
116 % compute trial von Mises true stress
117 q_tr = sqrt(3.*J_2);
118
119 % compute yield stress
120 fsig0 = sig_y0 + Rinf.*(1-exp(-gamma.*R0));
121
122 % evaluate yield function in terms of EFFECTIVE stress
123 Phi = q_tr./Int0 - fsig0;
124
125 %% Check if Yield Criterion is met
126 if Phi >= 0
127
128    %% return mapping initialization
129
130    % initial guess for the system variables
131    sig = sig_tr;
132    D = 1-Int0;
133    plasticMult = 0;
134    beta = beta0;
135
136    % initial values for iteration
137    norm_A = 1;
138
139    % initialization of Jacobian
140    dimJac = 2.*ncomp + 2;
141    jac = zeros(dimJac, dimJac);
142
143    % initialization of functions
144    A = zeros(4,1);
145
146    % initialization of internal variables
147    alpha_k = zeros(dimJac, 1);
148    alpha = alpha_k;
149
150    % jacobian locations
151    cs1 = 1;
152    ce1 = ncomp;
153    cs2 = ncomp+1;
```

```
154   cs3 = ncomp+2;
155   cs4 = ncomp+3;
156   ce4 = dimJac;
157
158   rs1 = 1;
159   rs2 = 2;
160   re2 = ncomp+1;
161   rs3 = ncomp+2;
162   re3 = dimJac-1;
163   rs4 = dimJac;
164
165   % inverse tangent
166   invCe = Ce\diag(ones(ncomp,1));
167
168   %% Newton-Raphson iteration for solution of non-linear coupled equations
169   % Orthogonal projection of the trial stress onto the yield surface
170   % Linearized problem is of the form:
171   % A(alpha) = 0 = A(alpha_k) + dA/dalpha * (alpha-alpha_k)
172   % Solution is:
173   % alpha = alpha_k - J^-1 A(alpha_k)
174   iter = 1;
175   maxiter = 20;
176   maxiter1 = 30;
177   maxiter2 = 30;
178   while norm_A >= tol && iter <= maxiter+maxiter2
179
180     if iter == maxiter+maxiter2
181       error('Fatal: Return mapping reached maximum iterations!');
182     end
183
184     %% reset jacobian and solution vectors
185     jac = zeros(dimJac, dimJac);
186     alpha_k = zeros(dimJac, 1);
187     alpha = zeros(dimJac, 1);
188
189     %% precalculations
190     % secondary system variables
191     R = R0 + plasticMult;
192     Int = 1-D;
193     sig_hyd = sum(sig.*Ih)./3;
194     sig_dev = sig - sig_hyd.*I;
195     rel = sig_dev - beta;
196     rel2 = rel .* physToEn;
197
198     % vMises stress and second invariant of relative stress
199     J_2 = sum((rel.^2).*physToEn)./2;
200     q = sqrt(3.*J_2);
201
202     % current yield stress, hardening modulus, flow vector, damage
203     % energy release rate, damaged elasticity tensor
204     fsig = sig_y0 + Rinf.*(1-exp(-gamma.*R));
205     dfsig = gamma .* Rinf .* exp(-gamma.*R);
206     N = 1.5 .* rel ./ (Int .* q);
207     N2 = 1.5 .* rel2 ./ (Int .* q);
208     Y = - (1./(2.*Int.^2)) .* sig' * invCe * sig;
```

```
209     DCe = Int .* Ce;
210
211     % some recurring factors
212     f1 = q.^2;
213     f3 = a .* plasticMult;
214     f4 = (- Y ./ r).^s;
215     f5 = Int.^2;
216     f6 = (-Y./r).^(s-1);
217     f7 = plasticMult .* s ./ (r.*Int);
218
219     % some recurring derivatives with respect to true stress
220     dq_dSig   = N .* Int;
221
222     % some recurring derivatives with respect to D
223     dSig_dD  = - sig ./ Int;
224     dRel_dD  = dSig_dD - (1./3) .* sum(dSig_dD.*Ih) .* I;
225     dJ2_dD   = sum((dRel_dD.*rel).*physToEn);
226     dq_dD    = .5 .* dJ2_dD .* sqrt(3./J_2);
227
228     %% compute all functions (this is A(alpha_k))
229     A(rs1)     = q ./ Int - fsig;
230     A(rs2:re2) = sig - DCe * (eps_e_tr - plasticMult .* N2);
231     A(rs3:re3) = beta - beta0 - plasticMult .* (a.*N - b.*beta);
232     A(rs4)     = D - D0 - (1./Int) .* f4 .* plasticMult;
233
234     % set the norm
235     norm_A     = norm(A);
236
237     %% compute all necessary derivatives
238
239     % dA1/d...
240     dA1_dSig    = N2;
241     dA1_dD      = q ./ f5 + dq_dD ./ Int;
242     dA1_dP      = -dfsig;
243     dA1_dBeta   = -N2;
244
245     % dN/d...
246     dN_dSig  = - 1.5 .* (rel*dq_dSig' - q .* Id1) ./ (Int .* f1);
247     dN_dD    = 1.5 .* (q .* (rel + Int.*dRel_dD) - Int.*dq_dD.*rel) ./ (f5.*f1);
248     dN_dBeta =   1.5 .* (rel*dq_dSig' - q.* Is1) ./ (Int .* f1);
249
250     % dA2/d...
251     dA2_dSig    = Is1 + plasticMult.*Int.*G2 .* dN_dSig;
252     dA2_dD      = dSig_dD + Ce * eps_e_tr + G2.*(Int .* plasticMult .* dN_dD - plasticMult.*N
                 );
253     dA2_dP      = Int.*G2 .* N;
254     dA2_dBeta   = plasticMult.*Int.*G2 .* dN_dBeta;
255
256     % dA3/d...
257     dA3_dSig    = - f3 .* dN_dSig;
258     dA3_dD      = - f3 .* dN_dD;
259     dA3_dP      = b .* beta - a .* N;
260     dA3_dBeta   = (1 + b .* plasticMult) .* Is1 - f3 .* dN_dBeta;
261
262     % dY/dSig
```

```
263     dY_dSig    = - (1./(Int.^2)) .* invCe*sig;
264     dA4_dSig   = f7 .* dY_dSig .* f6;
265     dA4_dD     = 1 - plasticMult .* f4 ./ f5;
266     dA4_dP     = - (1 ./ Int) .* f4;
267
268     %% fill Jacobian
269     jac(rs1, cs1:ce1)   = dA1_dSig';
270     jac(rs1, cs2)       = dA1_dD;
271     jac(rs1, cs3)       = dA1_dP;
272     jac(rs1, cs4:ce4)   = dA1_dBeta';
273
274     jac(rs2:re2, cs1:ce1) = dA2_dSig;
275     jac(rs2:re2, cs2)     = dA2_dD;
276     jac(rs2:re2, cs3)     = dA2_dP;
277     jac(rs2:re2, cs4:ce4) = dA2_dBeta;
278
279     jac(rs3:re3, cs1:ce1) = dA3_dSig;
280     jac(rs3:re3, cs2)     = dA3_dD;
281     jac(rs3:re3, cs3)     = dA3_dP;
282     jac(rs3:re3, cs4:ce4) = dA3_dBeta;
283
284     jac(rs4, cs1:ce1)   = dA4_dSig';
285     jac(rs4, cs2)       = dA4_dD;
286     jac(rs4, cs3)       = dA4_dP;
287
288     %% make a vector of the system variables (sigma, D, pM, beta)
289     alpha_k(cs1:ce1)  = sig;
290     alpha_k(cs2)      = D;
291     alpha_k(cs3)      = plasticMult;
292     alpha_k(cs4:ce4)  = beta;
293
294     %% solve the linear system
295     % alpha = alpha_k - J^-1 A(alpha_k)
296     alpha = alpha_k - jac\A;
297
298     %% update internals
299     sig        = alpha(cs1:ce1);
300     D          = alpha(cs2);
301     plasticMult = alpha(cs3);
302     beta       = alpha(cs4:ce4);
303
304
305     iter = iter + 1;
306 end % Newton Raphson Iteration
307
308
309 %% Consistent Elastoplastic Tangent Modulus
310
311 % compute all converged parameters for Jacobian
312 % as needed for the material tangent
313 % and the elastic strain
314
315 % secondary system variables
316 R = R0 + plasticMult;
317 Int = 1-D;
```

```
318   sig_hyd = sum(sig.*Ih)./3;
319   sig_dev = sig - sig_hyd.*I;
320   rel = sig_dev - beta;
321   rel2 = rel .* physToEn;
322
323   %% check if NR yielded an acceptable damage variable
324   if(Int < 10^-20)
325     error('Fatal: GP integrity too small!');
326   end
327
328   % vMises stress and second invariant of relative stress
329   J_2 = sum((rel.^2).*physToEn)./2;
330   q = sqrt(3.*J_2);
331
332   % current yield stress, hardening modulus, flow vector, damage
333   % energy release rate, damaged elasticity tensor
334   fsig = sig_y0 + Rinf.*(1-exp(-gamma.*R));
335   dfsig = gamma .* Rinf .* exp(-gamma.*R);
336   N = 1.5 .* rel ./ (Int .* q);
337   N2 = 1.5 .* rel2 ./ (Int .* q);
338   Y = - (1./(2.*Int.^2)) .* sig' * invCe * sig;
339   DCe = Int .* Ce;
340
341   % some recurring factors
342   f1 = q.^2;
343   f3 = a .* plasticMult;
344   f4 = (- Y ./ r).^s;
345   f5 = Int.^2;
346   f6 = (-Y./r).^(s-1);
347   f7 = plasticMult .* s ./ (r.*Int);
348
349   % some recurring derivatives with respect to sigma and beta
350   dq_dSig   = N.*Int;
351
352   % some recurring derivatives
353   dSig_dD   = - sig ./ Int;
354   dRel_dD   = dSig_dD - (1./3) .* sum(dSig_dD.*Ih) .* I;
355   dJ2_dD    = sum((dRel_dD.*rel).*physToEn);
356   dq_dD     = .5 .* dJ2_dD .* sqrt(3./J_2);
357
358   % reset jacobian
359   jac = zeros(dimJac, dimJac);
360
361   %% compute all necessary derivatives
362
363   % dA1/d...
364   dA1_dSig   = N2;
365   dA1_dD     = q ./ f5 + dq_dD ./ Int;
366   dA1_dP     = -dfsig;
367   dA1_dBeta  = -N2;
368
369   % dN/d...
370   dN_dSig   = - 1.5 .* (rel*dq_dSig' - q .* Id1) ./ (Int .* f1);
371   dN_dD     = 1.5 .* (q .* (rel + Int.*dRel_dD) - Int.*dq_dD.*rel) ./ (f5.*f1);
372   dN_dBeta  =  1.5 .* (rel*dq_dSig' - q.* Is1) ./ (Int .* f1);
```

```
373
374    % dA2/d...
375    dA2_dSig    = Is1 + plasticMult.*Int.*G2 .* dN_dSig;
376    dA2_dD      = dSig_dD + Ce * eps_e_tr + G2.*(Int .* plasticMult .* dN_dD - plasticMult.*N);
377    dA2_dP      = Int.*G2 .* N;
378    dA2_dBeta   = plasticMult.*Int.*G2 .* dN_dBeta;
379
380    % dA3/d...
381    dA3_dSig    = - f3 .* dN_dSig;
382    dA3_dD      = - f3 .* dN_dD;
383    dA3_dP      = b .* beta - a .* N;
384    dA3_dBeta   = (1 + b .* plasticMult) .* Is1 - f3 .* dN_dBeta;
385
386    % dA4/d...
387    dY_dSig     = - (1./(Int.^2)) .* invCe*sig;
388    dA4_dSig    = f7 .* dY_dSig .* f6;
389    dA4_dD      = 1 - plasticMult .* f4 ./ f5;
390    dA4_dP      = - (1 ./ Int) .* f4;
391
392    %% fill Jacobian
393    jac(rs1, cs1:ce1)   = dA1_dSig';
394    jac(rs1, cs2)       = dA1_dD;
395    jac(rs1, cs3)       = dA1_dP;
396    jac(rs1, cs4:ce4)   = dA1_dBeta';
397
398    jac(rs2:re2, cs1:ce1) = dA2_dSig;
399    jac(rs2:re2, cs2)     = dA2_dD;
400    jac(rs2:re2, cs3)     = dA2_dP;
401    jac(rs2:re2, cs4:ce4) = dA2_dBeta;
402
403    jac(rs3:re3, cs1:ce1) = dA3_dSig;
404    jac(rs3:re3, cs2)     = dA3_dD;
405    jac(rs3:re3, cs3)     = dA3_dP;
406    jac(rs3:re3, cs4:ce4) = dA3_dBeta;
407
408    jac(rs4, cs1:ce1)   = dA4_dSig';
409    jac(rs4, cs2)       = dA4_dD;
410    jac(rs4, cs3)       = dA4_dP;
411
412
413    ijac = jac\diag(ones(dimJac,1));
414
415
416    %% update model
417    % correction of trial state
418    % increase in plastic strain
419    deps_pl = plasticMult.*N2;
420    % true hydrostatic stress
421    sig_hyd = (1./3) .* sum(sig.*Ih);
422    % true stress
423    eldat.sigma(i, :,ig) = sig;
424    % plastic strain
425    eldat.eps_pl_n(i,:, ig) = eps_pl + deps_pl;
426    % damage
427    eldat.damage_n(i, ig) = D;
```

```
428    % back stress
429    eldat.beta_n(i, :, ig) = beta;
430    % isotropic hardening internal variable
431    eldat.R_n(i, ig) = R;
432    % true vM stress
433    eldat.q_n(i, ig) = q;
434    % stress triaxiality
435    eldat.triax_n(i, ig) = sig_hyd./q;
436    % yield stress
437    eldat.sigy_n(i, ig) = fsig;
438
439
440    % D
441    % fourth order elastoplastic tangent
442    % (minor symmetric, not major symmetric)
443
444    % from the Jacobian inverse we obtain the material
445    % tangent as a matrix with the upper corner at (1,2)
446    % and the lower corner at (ncomp,ncomp+1)
447    Dp = Int .* ijac(1:ncomp, 2:ncomp+1)*Ce;
448
449 % If yield criterion is not met, material is behaving elastically
450 else
451    % return damaged elastic tangent
452    Dp = Int0.*Ce;
453    % update model
454    eldat.sigma(i, :,ig) = sig_tr;
455    eldat.eps_pl_n(i,:, ig) = eps_pl;
456    eldat.damage_n(i, ig) = D0;
457    eldat.beta_n(i, :, ig) = beta0;
458    eldat.R_n(i, ig) = R0;
459    eldat.q_n(i, ig) = q_tr;
460    eldat.triax_n(i, ig) = sig_hyd_tr./q_tr;
461    eldat.sigy_n(i, ig) = fsig0;
462 end
```